\documentclass[10pt,twocolumn,twoside]{IEEEtran}
\usepackage{amsfonts,dsfont,amssymb,amsbsy,amsmath,paralist,theorem,bm,ifthen,color}
\usepackage[pdfstartview=FitH,bookmarksnumbered,unicode,bookmarksopen=true]{hyperref}
\usepackage{graphicx}
\usepackage{epstopdf}
\usepackage{multirow}
\usepackage{subfigure,relsize}
\usepackage{mathrsfs}
\usepackage{bm}
\usepackage{stfloats}
\usepackage{url}
\usepackage[noadjust]{cite}
\usepackage{cases,booktabs}
\usepackage{graphicx,algorithmic,algorithm,relsize}
\usepackage[justification=centering]{caption} 
\usepackage[table]{xcolor}

\newcommand\wb{\ensuremath{{\bf w}}}
\newcommand\yb{\ensuremath{{\bf y}}}

\renewcommand\sb{\ensuremath{{\bf s}}}
\newcommand\ub{\ensuremath{{\bm u}}}
\newcommand\Hb{\ensuremath{{\bf H}}}

\newcommand\ab{\ensuremath{{\bm a}}}

\newcommand\Fb{\ensuremath{{\bf F}}}

\newcommand\Gb{\ensuremath{{\bf G}}}

\newcommand\Ib{\ensuremath{{\bf I}}}

\newcommand\Pb{\ensuremath{{\bm P}}}

\newcommand\Tb{\ensuremath{{\bm T}}}
\newcommand\Xb{\ensuremath{{\bf X}}}
\newcommand\xb{\ensuremath{{\bf x}}}
\newcommand\Yb{\ensuremath{{\bf Y}}}
\newcommand\Ub{\ensuremath{{\bm U}}}
\newcommand\Rb{\ensuremath{{\bf R}}}

\newcommand\Nb{\ensuremath{{\bf N}}}
\newcommand\nb{\ensuremath{{\bf n}}}

\newcommand\Vb{\ensuremath{{\bf V}}}
\newcommand\vb{\ensuremath{{\bf v}}}
\newcommand\Wb{\ensuremath{{\bf W}}}
\newcommand\Zb{\ensuremath{{\bf Z}}}

\def\bX{{\boldsymbol{X}}}

\def\bI{{\boldsymbol{I}}}

\def\bg{{\boldsymbol{g}}}

\def\bw{{\boldsymbol{w}}}

\DeclareMathOperator{\trace}{Tr}

\newcommand\Ybs{\ensuremath{\boldsymbol{\mathcal{Y}}}}

\newcommand\Hbs{\ensuremath{\boldsymbol{\mathcal{H}}}}

\newcommand\Sbs{\ensuremath{\boldsymbol{\mathcal{S}}}}

\newcommand\E{\ensuremath{{\mathbb{E}}}}
\newcommand\blkdiag{\ensuremath{{\rm blkdiag}}}

\newcommand\tr{\ensuremath{{\rm Tr}}}

\newcommand\Cs{\ensuremath{{\mathbb{C}}}}
\newcommand\Rs{\ensuremath{{\mathbb{R}}}}

\newcommand\Nset  {\ensuremath{{\mathcal{N}}}}

\newcommand\st    {\ensuremath{{\rm s.t.}}}

\newcommand\Hf{\ensuremath{{\mathsf{H}}}}

\DeclareMathOperator*{\Tr}{Tr}
\graphicspath{{fig/}}

\definecolor{green}{RGB}{34	195	46}
\definecolor{red}{RGB}{220 0 0}

\usepackage{graphicx} 

\title{Distributed Signal Processing for Extremely Large-Scale Antenna Array Systems: State-of-the-Art and Future Directions}

\author{Yanqing Xu, \IEEEmembership{Member, IEEE,}
	Erik G. Larsson, \IEEEmembership{Fellow, IEEE,}
	Eduard A. Jorswieck, \IEEEmembership{Fellow, IEEE,} \\
	Xiao Li, \IEEEmembership{Senior Member, IEEE,} 
        Shi Jin, \IEEEmembership{Fellow, IEEE,}
        and Tsung-Hui Chang, \IEEEmembership{Fellow, IEEE}

		\thanks{T.-H. Chang is the corresponding author.}
        \thanks{Y. Xu, and T.-H. Chang are with the School of Science and Engineering, The Chinese University of Hong Kong, Shenzhen, and also with the Shenzhen Research Institute of Big Data, Shenzhen 518172, China (e-mail: xuyanqing@cuhk.edu.cn, tsunghui.chang@ieee.org).}
        \thanks{E. G. Larsson is with the Department of Electrical Engineering (ISY), Link{\"o}ping University, 581 83 Link{\"o}ping, Sweden (e-mail: erik.g.larsson@liu.se).}
        \thanks{E. A. Jorswieck is with the Institute for Communications Technology, Technical University of Braunschweig, 38106 Braunschweig, Germany (e-mail: e.jorswieck@tu-braunschweig.de).}
        \thanks{X. Li, and S. Jin are with the National Mobile Communication Research Laboratory, Southeast University, Nanjing 210096, China (e-mail: {li\_xiao, jinshi}@seu.edu.cn).}
        \thanks{This work has been submitted to the IEEE for possible publication.
        	Copyright may be transferred without notice, after which this version may
        	no longer be accessible.}
        }

\date{\today}

\begin{document}

\maketitle

\begin{abstract}
    Extremely large-scale antenna arrays (ELAA) play a critical role
    in enabling the functionalities of next generation
    wireless communication systems. However, as the number of antennas
    increases, ELAA systems face significant bottlenecks, such as
    excessive interconnection costs and high computational
    complexity. Efficient distributed signal processing (SP)
    algorithms show great promise in overcoming these challenges. In
    this paper, we provide a comprehensive overview of distributed SP
    algorithms for ELAA systems, tailored to address these
    bottlenecks. We start by presenting three representative forms of
    ELAA systems: single-base station ELAA systems, coordinated
    distributed antenna systems, and ELAA systems integrated with
    emerging technologies. For each form, we review the associated
    distributed SP algorithms in the literature. Additionally, we
    outline several important future research directions that are
    essential for improving the performance and practicality of ELAA
    systems.
\end{abstract}


\section{Introduction} 

\subsection{Multi-Antenna-Enabled Wireless Communications}
Wireless communications have undergone remarkable development over the
past few decades, driven by numerous technological
advancements. Innovations such as new air interfaces for accelerated
data transmission, advanced source and channel coding for improved
efficiency and reliability, and streamlined protocols to minimize
overheads have played pivotal roles. During this journey, the
invention of multi-antenna techniques stands out as one of the most significant milestones. In multi-antenna-enabled wireless communications, systems deploy multiple antennas at the
transmitter, the receiver, or both ends, allowing the exploitation of
additional spatial degrees of freedom. This provides substantial
benefits across three key areas: spatial diversity, spatial
multiplexing, and beamforming gain \cite{bjorson2023twenty}. By
harnessing these capabilities, multi-antenna technology substantially
boosts spectral and energy efficiency, enhances link reliability, and
delivers these improvements without the need for additional spectrum
\cite{larsson2014massive,lu2014overview,gershman2010convex,luo2010semidefinite}.

Multi-antenna techniques have been successfully used in modern
wireless communication systems, enhancing both commercial and private
networks worldwide. In mobile telecommunications, these technologies
are essential for $4$G long-term evolution (LTE), $5$G new radio (NR),
and the upcoming $6$G networks, facilitating substantial increases in
data rates and network capacity while significantly reducing latency
\cite{andrews2014will}. Additionally, multi-antenna systems play a
vital role in Wi-Fi technology, from the $802.11$n standard to the
latest $802.11$ax (Wi-Fi $6$), by supporting multi-user multiple-input
multiple-output (MU-MIMO)
\cite{perahia2013next,khorov2018tutorial}. This capability allows
routers to communicate with multiple devices simultaneously at the same radio resource, improving network throughput and
reducing latency in multi-device environments. Beyond terrestrial
networks, multi-antenna techniques are also critical in satellite
communications, where the link robustness and capacity can be enhanced
by directing signals more precisely towards specific areas on Earth
through advanced beamforming capabilities
\cite{you2020massive,kodheli2020satellite,heo2023mimo}. Moreover,
these antenna technologies are instrumental in specialized
applications such as radar and navigation systems used in the
aerospace and maritime sectors, enhancing safety in autonomous
vehicles and aircraft systems \cite{li2007mimo,talisa2016benefits}.

\subsection{Evolution of Multi-Antenna Techniques}
The benefits of using multiple antennas for communication were
observed at the very early stages of wireless communications
\cite{bjornson2023twenty}. However, the beamforming gains and spatial
diversity offered by multiple antennas were not thoroughly understood
until 1998 \cite{foschini1998limits}, which theoretically demonstrated
that MIMO systems could potentially increase the capacity of a
wireless point-to-point link linearly with the number of antennas
at the transmitter and receiver, without the need for
additional power or bandwidth. Subsequently, numerous fundamental
contributions have been made by the research community to understand
the performance limits of MIMO-based systems
\cite{telatar1999capacity,chuah2002capacity,alamouti1998simple,zheng2003diversity,love2003grassmannian,heath2001antenna,spencer2004zero,xu2017joint,tse2005fundamentals}. Among
these, the work \cite{telatar1999capacity} characterized the capacity
of multi-antenna channels under Gaussian channels, while the work
\cite{chuah2002capacity} extended these findings to more realistic
fading channels. A simple space-time coding technique was introduced
in \cite{alamouti1998simple} which enabled transmit diversity and laid
the foundation for later MIMO developments. Additionally, the
trade-off between diversity gain and multiplexing gain in MIMO systems
was explored in \cite{zheng2003diversity}. Parallel to these academic
advancements, MIMO technology was prepared for practical application
through standardization efforts by the 3rd generation partnership
project ($3$GPP)
\cite{3GPP_TS_36211,3GPP_TS_36212,3GPP_TS_36213,3GPP_TR_36913}. Thanks
to these theoretical contributions from the research community and
successful standardization, MIMO became one of the most important
enabling technologies in $4$G LTE systems.

\begin{figure*}[th]
	\begin{minipage}{0.99\linewidth}
		\centering
		\includegraphics[width=0.98\linewidth]{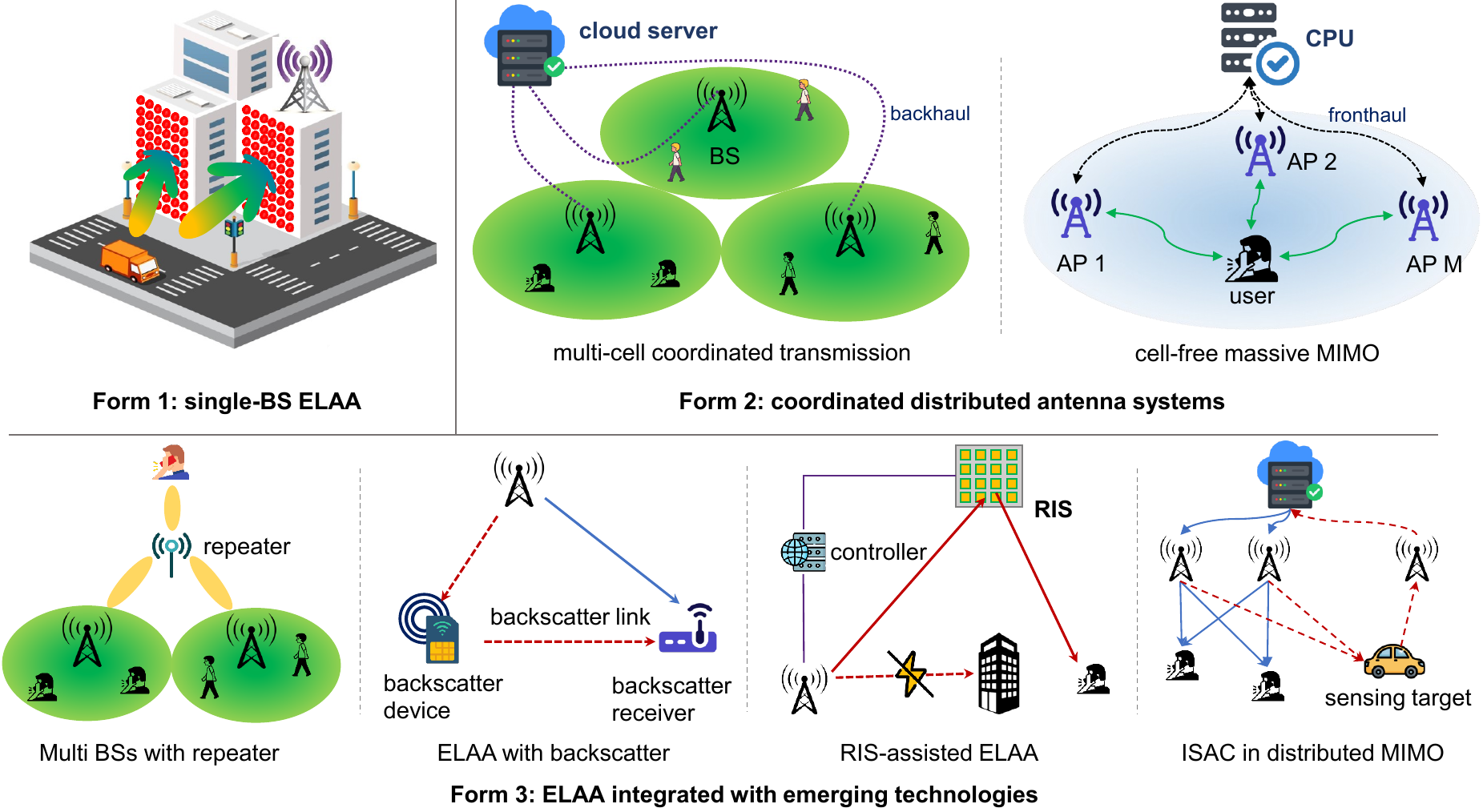}
		\caption{\small Illustrations of representative forms of ELAA systems (top row) and four strategies with relevant technologies (bottom row).}
		\label{fig: elaa}
	\end{minipage}
\end{figure*}

As an advancement of the MIMO technique, the concept of massive MIMO
was introduced in \cite{marzetta2010noncooperative}. In massive MIMO
systems, the number of antennas at the base station (BS) scales up by
orders of magnitude compared to traditional MIMO systems; for example,
configurations include $64$ or more antennas
\cite{rusek2012scaling,larsson2014massive,bjornson2016massive,sanguinetti2019toward}. Owing
to the massive number of antennas, these systems can leverage spatial
multiplexing to a much greater extent than was previously possible,
thereby achieving unprecedented improvements in throughput and
efficiency. Research indicates that by increasing the number of
antennas, massive MIMO systems using simple linear processing
techniques such as zero-forcing (ZF) and matched filtering (MF) can
achieve compelling spectral and energy efficiency
\cite{fundamentalsofmassivemimo,ho2017performance}. This shift not
only reduces the complexity but also increases the robustness of
wireless networks against fading and other channel impairments. Due to
these advantages, massive MIMO serves as a fundamental enabling
technique in $5$G NR to support enhanced mobile broadband, massive
machine-type communication, as well as ultra-reliable and low-latency
communication
\cite{Samsung_3GPP_RAN1_202024,3GPP_TR_38913,3GPP_TS_38214,wang2024two,jin2023massive}.

As we transition into the $6$G era, wireless systems are set to
deliver enhanced performance to enable much more sophisticated
applications such as immersive communication, massive communication, and hyper reliable and low-latency communication, as defined by the International Telecommunication Union
Radiocommunication Sector (ITU-R) \cite{ITU-R_2023}. Within this
framework, multi-antenna technologies continue to serve as a
cornerstone, evolving into increasingly advanced systems to meet these
ambitious goals. To achieve higher data rates, lower latency, and
greater connectivity envisioned for 6G, there is a pressing need to
advance antenna technologies by further increasing the number of antennas.
This has led to the emergence of extremely large-scale antenna array (ELAA) systems\footnote{
For some application scenarios, such as satellite communications, where adaptability and massive connectivities are not critical, one can use reflectarrays as a cost-effective approach to create large-scale antenna array systems \cite{zhu2022additively,mahouti2022computationally,mali2023design}.
}. The increase in the number of antennas can be achieved through two approaches: by augmenting the number of antennas at a single transmitter or by coordinating multiple transmitters for joint transmissions \cite{ngo2017cell,Amiri2018extremely,bjornson2020power,han2023towards,lu2024tutorial}.

With an extremely large number of antennas, ELAA systems offer significant advantages over conventional antenna configurations, including enhanced spatial resolution, better wireless channel sounding, and improved power efficiency, making them a promising solution for addressing the challenges of next-generation networks. 
The importance of ELAA as a critical enabling technology for next-generation networks has also been evidenced by industrial initiatives from leading telecom companies.  For instance, Nokia and AT\&T collaborated to demonstrate that the cooperative distributed massive MIMO system can increase the uplink capacity between $60\%$ and $90\%$ compared to similarly configured systems with a single panel, without sacrificing the downlink performance \cite{nokia_att_distributed_mimo}, and Huawei has also successfully deployed distributed massive MIMO systems in indoor cellular networks \cite{huawei_5g_distributed_mimo}.

\subsection{Features and Challenges of ELAA Systems}
ELAA systems can take various forms, each tailored to enhance wireless
network performance through advanced antenna technologies.  In this
paper, we particularly focus on three representative forms of ELAA
systems, as shown in Fig. \ref{fig: elaa}.  The first form is the
single-BS ELAA system, characterized by an extremely large number of
antennas at a single BS.  The second form involves coordinating a
large number of distributed antennas across multiple sites for joint
transmission. This approach leverages spatial diversity to a greater
extent, enhancing coverage and reducing signal fading and interference
across large areas.  The third form integrates ELAA systems with
emerging technologies, such as network-controlled repeaters,
reconfigurable intelligent surfaces (RIS), and wireless sensing nodes.
These technologies introduce additional flexibility and functionality,
enabling ELAA systems to adapt more effectively to dynamic
environmental conditions and user demands, thereby opening new avenues
for network optimization and service delivery.

The deployment of ELAA systems, while offering substantial benefits,
also introduces significant challenges due to the extremely large
number of antennas involved, including 1) excessive
interconnection cost between the antennas and baseband processor to
collect the received signals for processing, 2) high computational
complexity due to large-scale signal processing (SP) tasks, and 3)
complex synchronization and calibration across all antennas to fully
exploit the channel reciprocity for joint processing in time-division
duplex (TDD) systems.  These challenges present the main bottlenecks
in the practical implementation of ELAA systems. We will elaborate on
these challenges in detail in the next section. Addressing these
challenges is central to the discussions in this paper and will be
comprehensively presented in the subsequent sections, outlining both
current advancements and promising future research directions to
facilitate the practical deployment of ELAA systems.

In addition to the challenges discussed above, ELAA systems face an
additional challenge since they naturally operate in the radiating
near-field region due to the extremely large number of
antennas. Unlike in the far-field, the wireless channel in the
near-field region exhibits two distinct phenomena: 1)
\textit{spherical signal wavefronts}, where both the distance and the
angle between the source and each antenna array element differ
\cite{bjornson2020power}; 2) \textit{spatial nonstationarity}, where
different parts of the antenna array may observe the same channel
paths with varying power or even entirely different channel paths
\cite{Amiri2018extremely,gao2012measured,larsson2014massive}.  These
near-field phenomena require a different treatment compared to
classical wireless communication models that assume far-field plane
wave propagation. Moreover, the large physical size of ELAA 
systems and their operation in the near-field make them more susceptible 
to interference and in-site pattern distortions caused by nearby scatterers.
Despite posing challenges to SP algorithm design,
the near-field phenomena also offer several benefits, such as
increased spatial degrees-of-freedom and an improved ability to
enhance the signal strength of target receivers in wireless
communications. Interested readers may refer to
\cite{lu2024tutorial,han2023towards,wang2024tutorial} for recent
surveys on the near-field systems. To be complementary to
\cite{lu2024tutorial,han2023towards,wang2024tutorial}, in this
paper we primarily focus on the challenges of interconnection
cost, computational complexity, as well as synchronization and
calibration.

\subsection{Contributions and Organization of the Paper}
There exist several overview articles about large-scale antenna
systems that provide comprehensive insights into various aspects of
these systems. For instance, \cite{rusek2012scaling,lu2014overview}
offered fundamental overviews of the advantages and research
challenges associated with massive MIMO systems, discussing aspects
such as spatial multiplexing, energy efficiency, and hardware
implementation challenges. The papers
\cite{interdonato2019ubiquitous,elhoushy2021cell} focused on another
important enabling technique for ELAA, namely cell-free MIMO systems,
and highlighted the benefits of cell-free MIMO systems in providing
uniform service quality and mitigating inter-AP interference. Recent
works \cite{pan2022overview,zhang2020prospective} focused on
large-scale intelligent surfaces, such as intelligent reflecting
surfaces (IRSs) and reconfigurable intelligent surfaces (RISs). More
recently, near-field phenomena for ELAA systems were covered in
\cite{lu2024tutorial,han2023towards,wang2024tutorial}. Despite the
invaluable contributions provided by these studies, none of them
specifically focused on the distributed SP algorithm designs for ELAA systems, 
which are crucial for overcoming bottlenecks such as excessive 
interconnection costs and high computational complexity.

The goal of this paper is to provide a comprehensive survey of 
distributed SP algorithms tailored for three representative forms 
of ELAA systems. Specifically, we focus on distributed SP algorithm 
designs to address the challenges of interconnection cost and 
computational complexity encountered in ELAA systems. This emphasis 
ensures that this paper stands apart from existing works 
\cite{rusek2012scaling, lu2014overview, interdonato2019ubiquitous, elhoushy2021cell, pan2022overview, zhang2020prospective}, while being complementary to \cite{lu2024tutorial, han2023towards, wang2024tutorial}, as none of them specifically 
focus on the distributed SP algorithm designs for ELAA systems.
The detailed structure of the rest of this paper is as follows.
\begin{itemize}
    \item Section \ref{sec: elaa in focus} introduces the three representative 
    forms of ELAA systems and outlines the research challenges associated with them.
    It clarifies that efficient distributed SP algorithms, based on the decentralized 
    baseband processing architecture of ELAA systems, are essential for 
    overcoming these challenges. 
    \item Section \ref{sec: single-BS} reviews the distributed SP algorithms 
    for single-BS ELAA systems using DBP architectures. 
    We first introduce the system model and signal model, 
    then review the distributed channel estimation, 
    distributed uplink equalization, and distributed downlink precoding algorithms. 
    Finally, we discuss the potential errors in channel state information (CSI) 
    and their impact on system performance.
    \item Section \ref{sec: distributed antenna} provides an overview 
    of distributed SP algorithms for coordinated distributed antenna systems, 
    with a focus on the coordinated multi-cell transmission and CF-mMIMO scenarios.
    \item Section \ref{sec: emerging technology} explores the interactions between emerging technologies and ELAA systems. In particular, we consider four key technologies: 
    network-controlled repeaters, backscatter communication, RIS, 
    and integrated sensing and communication (ISAC).
    \item Section \ref{sec: outlook} outlines several future research directions, 
    including distributed SP algorithms for ELAA systems in the near-field, ELAA with flexible antennas, ELAA for physical layer security, low-resolution designs for ELAA systems, and ELAA for satellite communications.
    \item Section \ref{sec: conclusion} concludes the paper.
\end{itemize}

\textbf{Notation:} Column vectors and matrices are denoted by
boldfaced lowercase and uppercase letters, e.g., $\xb$ and $\Xb$.
$\mathbb{C}^{n \times n}$ stands for the sets of $n$-dimensional
complex matrices.  The superscripts $(\cdot)^\top$ and $(\cdot)^\Hf$
describe the transpose and Hermitian operations, respectively.
$\tr(\bf X)$ represents the trace of $\Xb$.  $||\xb||^2$ and
$||\Xb||^2$ denote squares of the Euclidean norm and Frobenius norm of
vector ${\xb}$ and matrix $\Xb$, respectively.  $\mathcal{R}(\Xb)$
returns the real part of a complex-valued matrix $\Xb$.  $\E\{\cdot\}$
represents the statistical expectation operation. $\odot$ denotes the
Hadamard (element-wise) product.

\section{ELAA in Focus: Representative Forms and Research Challenges} \label{sec: elaa in focus}
In this section, we provide a comprehensive introduction to ELAA
systems. We begin by introducing three typical forms of ELAA
systems. We then discuss the significant challenges associated with
these systems, primarily stemming from the use of an extremely large
number of antennas. Lastly, we introduce the DBP architecture, which
is acknowledged as a promising approach to overcoming these
challenges.

\subsection{Three Forms of ELAA Systems}

As shown in Fig. \ref{fig: elaa}, the ELAA systems can appear in the following three forms:
\begin{enumerate}
	\item {\bf Single-BS ELAA system:} To enhance wireless
          communication system performance, a natural evolution of
          contemporary massive MIMO technology involves substantially
          increasing the number of antennas at the BS. Typically, this
          increase is by an order of magnitude, ranging from several
          hundreds to thousands of antennas, resulting in the ELAA
          system
          \cite{rusek2012scaling,larsson2014massive,marzetta2010noncooperative,lu2024tutorial,han2023towards,bjornson2020power}. Thanks
          to the unprecedented improvement in spatial resolution
          brought about by the extremely large number of antennas, the
          ELAA system is regarded as a key enabling technology for 6G,
          meeting several stringent key performance indicators such as
          spectral efficiency, energy efficiency, reliability, as well
          as positioning and sensing accuracy.

    \item {\bf Coordinated distributed antenna systems:} Instead of
      directly increasing the number of antennas at a BS,
      another approach to implementing ELAA systems is to coordinate a
      large network of distributed antennas for joint
      transmission. Two prominent examples of such systems are the
      coordinated multi-cell (joint) transmission system
      \cite{bjornson2010cooperative,bhagavatula2011adaptive} and the
      CF-mMIMO system \cite{ngo2017cell}. In particular, the
      coordinated multi-cell transmission system involves a central
      server that communicates with multiple BSs via backhaul links to
      facilitate coordinated transmission, which has evolved into what
      is known as multi-radio multi-connectivity in the 5G
      New Radio standard \cite{3gpp2023mrdc}. Meanwhile, in the
      CF-mMIMO system, multiple APs are orchestrated by a CPU through
      fronthaul links to jointly serve multiple users efficiently.
	
    \item {\bf ELAA system integrated with emerging technologies:}
      Recently, a variety of emerging communication technologies have
      been integrated into large-scale antenna systems to enhance
      performance and introduce new functionalities while maintaining
      manageable complexity. As illustrated in Fig. \ref{fig: elaa},
      we are particularly interested in four technologies:
      network-controlled repeaters, backscatter communication
      techniques, RIS, and ISAC techniques.
\end{enumerate}

\subsection{Challenges of ELAA Systems}
To fully exploit the spatial degrees of freedom offered by ELAA
systems, most existing transceiver signal processing (SP) algorithms
are designed for centralized implementation, relying on a centralized
baseband processing (CBP) architecture, particularly in the single-BS
scenario. Such CBP architecture has been widely used in recent
multi-user massive MIMO testbed implementations, such as the Argos
testbed \cite{shepard2012argos,shepard2013argosv2}, the LuMaMi testbed
\cite{vieira2014a,malkowsky2017the}, and the BigStation testbed
\cite{yang2013bigstation}.  In the CBP architecture, signals from all
antennas are pooled and processed in a centralized baseband processor,
e.g., the cloud server or the CPU. However, as the number of antennas
grows, these centralized algorithms begin to face the following
challenges:
\begin{figure}[t]
	\centering
	\subfigure[Fronthaul costs]
	{\includegraphics[width=0.82\linewidth]{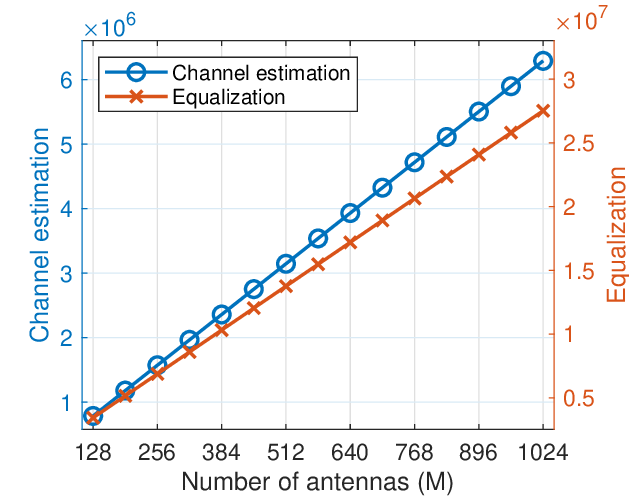}\label{fig: fronthaul cost}}\\
	\subfigure[Computational complexities]{
		\includegraphics[width=0.82\linewidth]{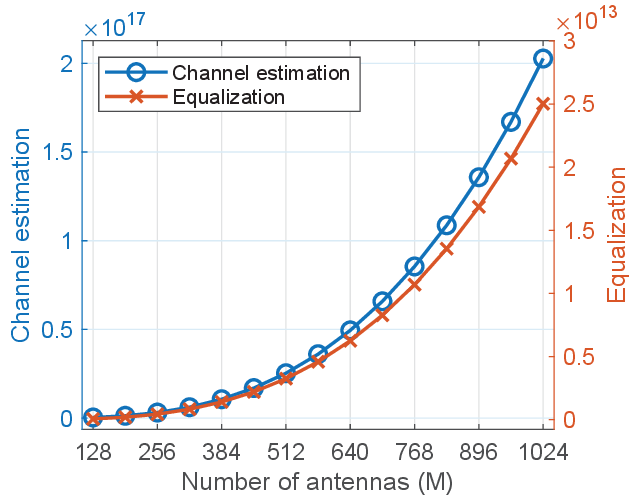}\label{fig: complexity}}
        \captionsetup{justification=justified, singlelinecheck=false, font=small}	
	\caption{\small The fronthaul costs (Fig. \ref{fig: fronthaul cost}) and computational complexities (Fig. \ref{fig: complexity}) of the centralized channel estimation algorithm and equalization algorithm based on the LMMSE criteria, where we use a setting of $192$ subcarriers (i.e., $16$ resource blocks), $32$ users, and $140$ symbols.}
	\label{fig : distributed network}
\end{figure}
\begin{itemize}
    \item {\bf Excessive interconnection cost}: The rapid growth in
      the number of BS antennas generates a substantial amount of raw
      baseband data, including CSI, transmit and received signals, which need to be
      exchanged between the CBP unit and distributed units. This
      results in a high communication burden and network delays when
      the interconnection bandwidth is limited. As an example, Fig. \ref{fig: fronthaul cost} illustrates the fronthaul costs of the centralized channel estimation and uplink equalization versus number of antennas. The fronthaul cost is measured by the number of complex numbers sent from the antennas to the central processor. As shown, the fronthaul cost increases linearly with the number of BS antennas.
      Another analysis conducted by the work \cite{li2017decentralized} show that the raw
      baseband data throughput exceeds $1$ Tbps for a massive MIMO BS
      operating at $80$ MHz bandwidth with $256$ BS antennas and
      $12$-bit digital-to-analog converters. Such demanding bandwidth
      requirement severely exceeds current BS internal interface
      standards such as the enhanced common public radio interface \cite{ecpri}. 
      
    \item{\bf High computational complexity}: The large number of
      antennas also results in high-dimensional signal processing
      problems. For example, the conventional linear minimum mean
      square error (LMMSE) estimator, commonly employed for channel
      estimation and equalization, requires high-dimensional matrix
      inversion operations. The computational complexity of these
      operations scales cubically with the number of BS antennas in practical scenarios with cross-cell interference, making it increasingly impractical for ELAA systems where the
      complexity becomes prohibitively expensive. This challenge is illustrated in Fig. \ref{fig: complexity}, which simulates the computational complexities of centralized channel estimation and uplink equalization with different number of antennas based on the LMMSE criterion.

    \item {\bf Complex synchronization and calibration:} Relying on
      channel reciprocity in time division duplex (TDD) systems offers
      an efficient way to simplify downlink channel estimation in ELAA
      systems. However, due to the inherent non-reciprocity of
      transceiver hardware, synchronization and calibration are
      essential to effectively exploit channel reciprocity in
      practice. Owing to the extremely large number of antennas,
      achieving high-precision synchronization and calibration can be
      even more challenging.
\end{itemize}
The combination of these challenges represents a major bottleneck in
ELAA systems, not only degrading system efficiency but also imposing
significant limitations on scalability.

\subsection{ELAA with Decentralized Baseband Processing}

It is noted that in the coordinated distributed antenna systems and
the ELAA system integrated with emerging technologies, each BS/AP is
equipped with a local baseband unit (BBU) which can be exploited for
processing locally received signals.  This means that, although many of
the existing algorithms are designed for centralized implementation in
these systems, they also inherently possess a DBP
architecture. Compared to CBP architecture, the DBP architecture has
the following advantages. First, a DBP architecture allows the
distributed nodes (DNs) to exchange some locally processed
(low-dimensional) intermediate results, thereby reducing the
interconnection costs. Second, since each DN only needs to process a
low-dimensional received signal, the computational complexity in each
DN can be significantly reduced. Last but not least, the DBP
architecture improves the scalability and robustness of ELAA systems,
as adding or removing a BS/AP simply amounts to adding or removing a
computing unit. Therefore, one can leverage the DBP architecture to
overcome the bottlenecks of excessive interconnection cost and high
computational complexity by developing efficient
distributed/decentralized SP algorithms.

Recently, the DBP architecture has also been introduced to the single-BS
ELAA system \cite{li2017decentralized}. As depicted in Fig. \ref{fig:
  dbp arch}, the antennas are grouped into several non-overlapping
clusters, with each cluster equipped with an independent and
cost-effective BBU. Each antenna cluster, in conjunction with its
local BBU, forms a DN which functions analogously
to a BS or an AP in distributed antenna systems. These DNs are
interconnected using various network topologies, such as the star
topology and the daisy-chain topology, through fronthaul links. This
architecture enables DNs to communicate and coordinate for joint
transmission, while simultaneously maintaining low complexity through
the use of efficient distributed SP algorithms.

Although the DBP architecture presents a promising solution to challenges such as excessive interconnection costs and high computational complexity faced by conventional CBP architectures, its effectiveness relies heavily on the development of efficient distributed SP algorithms. In the following sections, we provide a comprehensive overview of distributed SP algorithms for ELAA systems, specifically designed to address these critical bottlenecks.

\section{Distributed SP for Single-BS ELAA Systems} \label{sec: single-BS}

In this section, we delve into the intricacies of distributed SP
algorithms tailored for single-BS ELAA systems employing the DBP
architecture. We begin by introducing the system model of ELAA with
DBP, laying the foundation for understanding the distributed SP
framework. Next, we provide a comprehensive overview of key signal
processing tasks, encompassing channel estimation, uplink
equalization, and downlink precoding, which are crucial for optimizing
performance in ELAA systems. Finally, we focus on the synchronization
and calibration algorithms, pivotal for achieving coherent distributed
processing in ELAA systems.

\begin{figure}[t]
	\centering
	\includegraphics[width=0.78\linewidth]{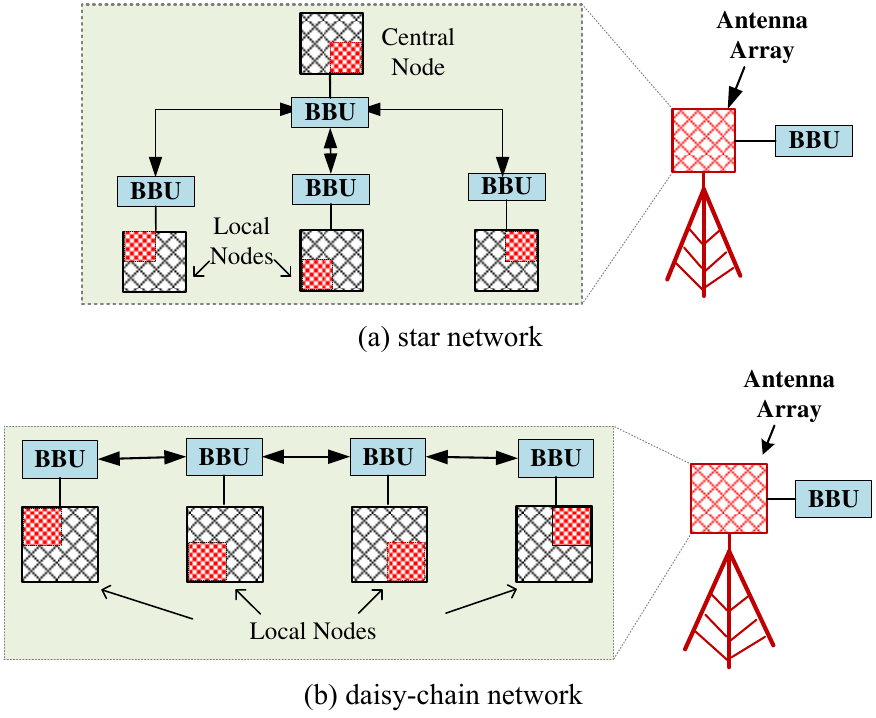}
        \captionsetup{justification=justified, singlelinecheck=false, font=small}	
	\caption{\small Single-BS ELAA system with antenna clustering-based DBP architecture.} 
	\label{fig: dbp arch}
\end{figure}

\subsection{System Model}
Consider a multi-carrier ELAA system where a single BS
equipped with $M$ antennas serves $L$ single-antenna users over
$N_{sc}$ subcarriers. We assume that the antennas within a single panel are spaced at half-wavelength ($\lambda/2$) apart, which is a typical setting in contemporary massive MIMO systems.
The antennas of the BS are divided into $C$
non-overlapping antenna clusters, with each cluster consisting of
$M_i$ antennas, where $\sum_{i=1}^C M_i = M$, and connecting to an
independent BBU to handle its received signal.  For ease of notation,
we refer to an antenna cluster together with its BBU as a DN.  Each
performs local signal processing, including channel estimation, uplink
equalization, and downlink precoding. The DNs are coordinated to
improve the system performance.  According to the way of information
exchange, the DNs connect as different distributed architectures,
e.g., the star network or the daisy-chain network, as shown in
Fig. \ref{fig: dbp arch}.  Based on this model, we overview the DSP
algorithm designs for ELAA using DBP architecture in the following
subsections.

\subsection{Channel Estimation}
Consider an orthogonal case and focus on the channel estimation of one
user. The antenna-and-frequency domain received pilot signal at DN $i$
is given by
\begin{align}
	\Yb_i = \Hb_i \Pb + \Nb_i \in \Cs^{M_i\times N_{sc}}, i \in \mathcal{C} \triangleq \{1,...,C\},
 \label{eq:1}
\end{align}
where $\Hb_i \in \Cs^{M_i \times N_{sc}}$ is the antenna-and-frequency
domain channel between the user and DN $i$, $\Nb_i \in \Cs^{M_i \times
  N_{sc}}$ is the received noise at DN $i$, $\Pb \in \Cs^{N_{sc}
  \times N_{sc}}$ denotes the pilot matrix.  The idea of distributed
channel estimation (DCE) is that each DN exploits the local received
signal $\Yb_i$ and coordinates with the other DNs to approach the
performance of the centralized CE algorithm.

There are only a few works in the literature that have studied the DCE
problem
\cite{Zaib2016Distributed,Zuo2021distributed,trigka2021distributed,tan2024federated,xu2023low}. Specifically,
\cite{Zaib2016Distributed} assumes that the antennas are deployed in a
uniform planar array, with each antenna equipped with a BBU that can
communicate only with its neighbors. Under this setting, an iterative
DCE algorithm was proposed.  However, this algorithm does not exploit
channel sparsity in the angle and delay domains. To improve estimation
accuracy, angle-domain channel sparsity was exploited in
\cite{Zuo2021distributed,trigka2021distributed}, where two distributed
algorithms based on an accelerated projection-based consensus method
and the orthogonal matching pursuit method were proposed,
respectively, to address the sparse channel estimation
problem. Nonetheless, all the proposed DCE algorithms in
\cite{Zaib2016Distributed,Zuo2021distributed,trigka2021distributed}
are iterative in nature, which leads to significant processing delays
due to frequent information exchanges among the DNs. Additionally, the
computational complexity linearly increases with the number of
iterations. Furthermore, channel sparsity in the delay domain was not
fully exploited in these works, thus suffering a further performance
loss. Recently, to overcome these disadvantages, \cite{xu2023low}
proposed a low-complexity sparse aggregation-based DCE framework,
which can perform as well as centralized schemes while maintaining low
fronthaul costs and a computational complexities.  In the next
subsection, we will elaborate on this DCE framework in detail.

\subsubsection{Sparse Aggregation-Based DCE Framework}
In \cite{xu2023low}, the low-complexity diagonal MMSE (DMMSE)
estimator is considered, while the developed DCE framework applies to
other powerful estimators, e.g., the LMMSE estimator
\cite{chang2010training,takano2018conditional} and compressed
sensing-based estimators \cite{wei2017near,cheng2019adaptive}.  By
concatenating the received signals, channel matrices, and noise
matrices, we have {\small $\Yb \!=\!
  [\Yb_1^{\Hf},\Yb_2^{\Hf},\ldots,\Yb_C^{\Hf}]^{\Hf}$, $\Hb \!=\!
  [\Hb_1^\Hf,\Hb_2^\Hf,\ldots,\Hb_C^\Hf]^\Hf$}, and {\small $\Nb \!=\!
  [\Nb_1^\Hf,\Nb_2^\Hf,\ldots,\Nb_C^\Hf]^\Hf$}.  Here, for simplicity,
we assume that the pilot matrix is an identical matrix.  Then, based
on the DMMSE estimator \cite{shariati2014low}, the centralized channel
estimation is to solve
\begin{align} \label{eqn: d-mmse estimator}
    \min_{\Sbs} \E\{\|\Sbs\odot \Ybs - \Hbs\|_F^2\},
\end{align}
where {\small $\Ybs = \Fb_M^\Hf \Yb \Fb_{N_{sc}}^\Hf$} and {\small
  $\Hbs = \Fb_M^\Hf \Hb \Fb_{N_{sc}}^\Hf$} are the angle-and-delay
domain received signal and channel matrix, respectively, with {\small
  $\Fb_M \in \Cs^{M\times M}$} and {\small $\Fb_{N_{sc}} \in
  \Cs^{N_{sc} \times N_{sc}}$} denoting the associated discrete
Fourier transformation (DFT) matrices; $\Sbs \in \Cs^{M \times
  N_{sc}}$ is the DMMSE estimator, and `$\odot$' signifies the
Hadamard (element-wise) product operator.  It is noted that problem
\eqref{eqn: d-mmse estimator} focuses on the angle-and-delay domain
channel estimation for two primary reasons.  Firstly, unlike the LMMSE
estimator which is equivalent when being applied in the
angle-and-delay domain and in the antenna-and-frequency domain, the
DMMSE estimator is preferable in the angle-and-delay domain since it
can achieve a lower mean square error (MSE) performance \cite[Theorem
  1]{xu2023low}.  Secondly, as will be demonstrated, estimating the
channel from the angle and delay domains is essential for reducing
both fronthaul costs and computational complexity in the DCE
algorithm.

The DMMSE estimator in problem \eqref{eqn: d-mmse estimator} actually
performs entry-wise channel estimation, and the $(m,j)$-th entry of
the DMMSE estimator {\small $\Sbs$} is given by {\small $[\Sbs]_{m,j}
  = \frac{[\Rb_{\Hbs}]_{m,j}}{[\Rb_{\Hbs}]_{m,j} + \sigma_w^2}, m \in
  \mathcal{M} \triangleq \{1,...,M\}, j \in \mathcal{N}_{sc}
  \triangleq \{1,...,N_{sc}\}$}. Here, {\small $\Rb_{\Hbs} \triangleq
  \E\{\Hbs \odot \Hbs^*\} \in \Rs^{M \times N_{sc}}$} is the power
profile of {\small $\Hbs$}.  Then, the centralized angle-and-delay
domain channel estimate can be written as
\begin{align} \label{eqn: central ce}
    \widehat\Hbs = \Sbs \odot \bigg(\sum_{i=1}^C\Fb_i^\Hf \Yb_i \Fb_{N_{sc}}^\Hf\bigg),
\end{align}
where $\Fb_i \in \Cs^{M \times M_i}, i \in \mathcal{C}$, is the
$i$-th submatrix of $\Fb_{M}$ by partitioning it horizontally into $C$
submatrices, i.e., $\Fb_M=[\Fb_1^\Hf, \Fb_2^\Hf,
  \ldots,\Fb_C^\Hf]^\Hf$.

Given the decomposable structure in \eqref{eqn: central ce} over DNs
and exploiting the angle-and-delay domain sparsity of the wireless
channel, \cite{xu2023low} proposed a low-complexity DCE framework
based on two different strategies -- first aggregating information
from DNs followed by estimating the channel, or first estimating the
channels at DNs followed by aggregating the estimates. These two
strategies lead to two low-complexity DCE algorithms, namely, the
aggregate-then-estimate (AGE) based and estimate-then-aggregate (EAG)
based algorithms.  In what follows, we give a brief introduction to
the AGE-based algorithm in the star-networked topology shown in
Fig. \ref{fig: dbp arch}(a). Specifically, the AGE-based algorithm
consists of the following three steps:

{\bf \indent i) Processing at DNs: } Each DN first performs an inverse
discrete Fourier transform (IDFT) operation to transform the local
received antenna-and-frequency domain signals into sparse
antenna-and-delay domain signals. Then, each DN performs a local
windowing operation on the local antenna-and-delay domain signals to
divide them into two parts. The part with larger values will be sent
to the CN for centralized channel estimation, while the part with
smaller values will be used for local channel estimation at each DN.

{\bf \indent ii) Processing at CN:} After receiving the sparse
antenna-and-delay domain signals from the DNs, the CU performs an IDFT
operation on each antenna-and-delay domain signal with the
corresponding $\Fb_i$ in \eqref{eqn: central ce} to transform them
into the angle-and-delay domain. Then, the angle-and-delay domain
signals of all DNs are aggregated by summing them together. The
centralized channel estimation is performed on the aggregated
angle-and-delay domain signal. After that, the CN performs a DFT
operation on the centralized estimate to transform it into
antenna-and-delay domain channel estimates and sends them to the
corresponding DNs.

{\bf \indent iii) Post-processing at DNs: } Each DN combines the
centralized channel estimate and the local channel estimate to obtain
the complete local antenna-and-delay domain channel estimate. Finally,
the antenna-and-frequency domain channel estimate of each DN is
obtained by performing a DFT operation on the local antenna-and-delay
domain channel estimate.

\begin{figure}[t]
	\centering
	\includegraphics[width=0.92\linewidth]{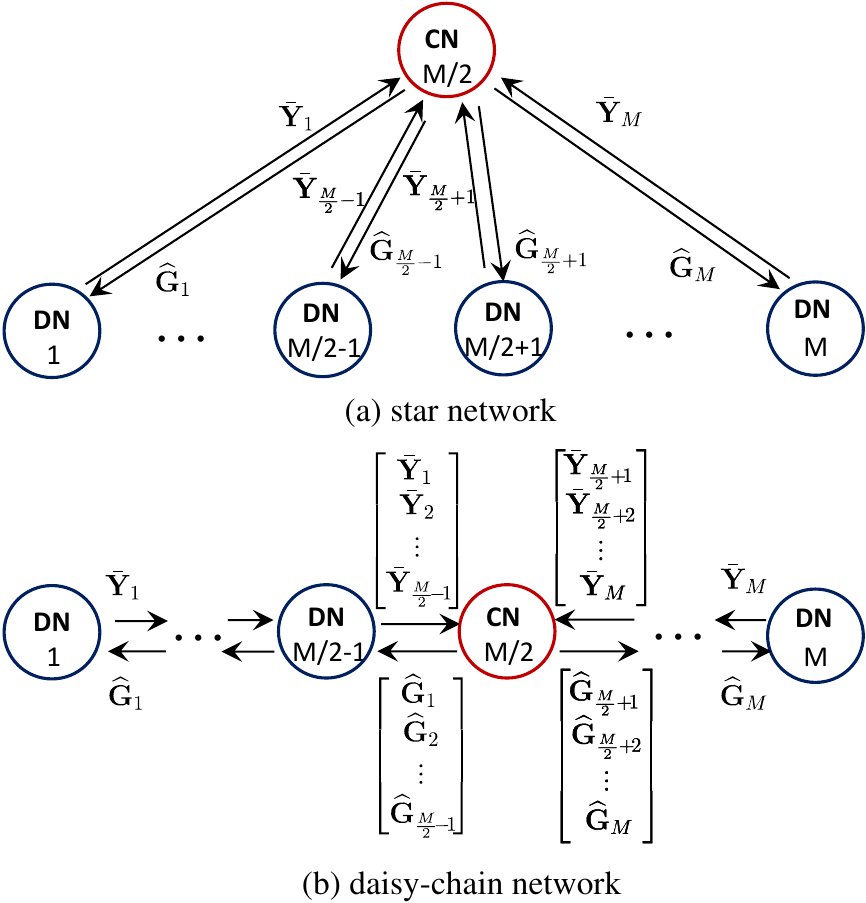}\\
        \captionsetup{justification=justified, singlelinecheck=false, font=small}
	\caption{\small Illustration of the signal exchange processes of the AGE-based algorithm in (a) the star network and (b) the daisy-chain network.} 
	\label{fig:information_exchange_stationary_dl_agg}
\end{figure} 

The AGE-based algorithm can also be directly applied to the
daisy-chain network. The only difference is how the signals are
exchanged among nodes. The detailed signal exchange processes of
AGE-based algorithm in the two networks are given in
Fig. \ref{fig:information_exchange_stationary_dl_agg}.

The proposed EAG-based algorithm in \cite{xu2023low} performs similar
to the AGE-based algorithm. Specifically, by the EAG-based algorithm,
each DN first estimates its local CSI in the angle-and-delay domain
and then sends them to the CN for aggregation and refined estimation.
Since both the angle- and delay-domain sparsities are exploited, the
EAG-based algorithm can achieve a similar performance as the AGE-based
algorithm but with an even smaller fronthaul cost. As a tradeoff, the
computational complexity of the EAG-based algorithm is slightly higher
than the AGE-based algorithm due to the refined estimation at the CN.

Compared to the DCE algorithms in
\cite{Zaib2016Distributed,Zuo2021distributed,trigka2021distributed},
the primary merits of the AGE- and EAG-based algorithms are that they
only require one roundtrip exchange information between
BBUs. As a result, the computational complexity and fronthaul cost are
significantly reduced.  In addition, AGE- and EAG-based algorithms
exploit the channel sparsity in the angle and delay domains and allow
flexible control of the tradeoff between fronthaul cost and estimation
accuracy. Numerical results in \cite{xu2023low} show that the overall
computational complexities of both algorithms are far smaller than
that of the centralized scheme, and the computation loads are evenly
distributed among all BBUs. This implies that these DCE algorithms are scalable with an increasing number of DNs.
Besides, the proposed algorithms can
achieve a comparable MSE performance as the centralized scheme with
significantly reduced fronthaul costs in both low and high
signal-to-noise ratio (SNR) regimes.

\subsubsection{Future Work}
One crucial direction for future research in channel estimation for
ELAA systems involves addressing the pilot contamination problem. The
existing DCE algorithms have assumed that different users are
scheduled on different pilot signals to avoid interference. However,
the future of wireless communication aims to support a significantly
larger number of users, while the number of available pilot signals
remains limited. This limitation results in the pilot contamination problem, where the reuse of pilot signals among different users introduces interference and significantly reduces channel estimation accuracy. Thus, developing DCE algorithms capable of effectively mitigating pilot contamination is crucial for enhancing the performance of ELAA systems.

\subsection{Uplink Equalization}
Advanced multiuser equalization (MUE) techniques capable of leveraging
the gain offered by the massive antennas play an important role in SP
algorithm designs for ELAA systems.  For uplink equalization, the
received signal at DN $i$ on subcarrier $j$ is given by
\begin{align}
	\yb_{i,j} = \Hb_{i,j} \sb_j + \ub_{i,j}, i \in \mathcal{C}, j \in \mathcal{N}_{sc},
\end{align}
where $\Hb_{i,j} \in \Cs^{M_i \times L}$ is the channel between DN $i$
and the $L$ users, $\sb_j \in \Cs^{L \times 1}$ denotes the users'
signal, and $\ub_j \in \Cs^{M_i \times 1}$ is the (colored or
white) noise.  In the literature, the LMMSE-based equalizer is widely
used owing to its low computational complexity compared to the
nonlinear maximum-likelihood detector and its high performance
compared to the linear counterparts, such as the maximum ratio
combining (MRC) and zero-forcing (ZF) based detectors
\cite{rusek2012scaling}.  By letting $\yb_j =
     [\yb_{1,j}^\top,\yb_{2,j}^\top,...,\yb_{C,j}^\top]^\top$ and
     based on the LMMSE criteria, the centralized equalization problem
     can be formulated as
\begin{align}  \label{eqn: lmmse equalization problem}
	\min_{\Wb_j} ~\mathbb{E}\left[\left\|\sb_j - \Wb_j
	\yb_j \right\|^{2}\right], \forall j \in  \Nset,
\end{align}
where $\Wb_j = [\Wb_{1,j},\Wb_{2,j},...,\Wb_{C,j}]$ is the equalizer
with $\Wb_{i,j} \in \Cs^{L \times M_i}$ representing the equalizer for
DN $i$.

Based on the decomposable structure of \eqref{eqn: lmmse equalization
  problem} over DNs, several works have studied the
distributed/decentralized equalization algorithm designs in the
literature
\cite{jeon2017achievable,zhang2020decentralized,dong2022enhanced,li2022gaussian,zhang2022decentralized,jeon2019decentralized,croisfelt2021decentralized,kulkarni2021hardware,rusek2022distributed,amiri2022uncoordinated}.
In particular, considering a star or a daisy-chain network, the works
\cite{jeon2017achievable,zhang2020decentralized,dong2022enhanced,li2022gaussian,zhang2022decentralized}
investigated the uplink MUE problem based on the iterative approximate
message passing algorithm, while the MRC and LMMSE-based MUE methods
are studied in \cite{jeon2019decentralized}.  Recently, considering an
extremely large-scale MIMO system, \cite{croisfelt2021decentralized}
studied the decentralized MUE problem under the spatially
non-stationary channel model.  While most existing works considered
the star or daisy-chain topologies, \cite{kulkarni2021hardware}
investigated the decentralized equalization in a ring topology under
the DBP architecture.  Despite these promising improvements, the
previous works all have assumed that the additive noise is spatially
white while ignoring the inter-cell interference.  Besides, these
algorithms require iterative information exchanges between distributed
nodes, which could incur a large amount of fronthaul cost. Therefore,
these algorithms maynot be applicable to the practical systems where
the inter-cell interference exists and the fronthaul bandwidth is
limited.

On the other hand, the number of antennas in each DN could be much
larger than the number of users in massive MIMO systems, indicating
that the received signal at each DN contains considerable redundancy.
This observation motivates compression-based equalization schemes
under the DBP architecture, where the DNs first compress the local
received signal to remove the redundancy and then forward the
(low-dimensional) compressed signal to the CN for joint equalization.
This compression-and-forward strategy has been widely studied for
bandwidth-limited systems, for example, the fronthaul/backhaul
capacity-limited cloud radio access network (C-RAN)
\cite{park2013joint,zhou2014optimized,park2014fronthaul,dai2016energy,zhou2016fronthaul,wiffen2021distributed},
where the compression is implemented by using sophisticated source
coding methods, e.g., the Wyner-Ziv coding in
\cite{park2013joint,zhou2014optimized,zhou2016fronthaul}.  As an
important branch of compression schemes, linear compression (LC) has
been investigated in sensor networks for distributed parameter
compression-estimation, in view of its low complexity
\cite{song2005sensors,schizas2007distributed,zhang2019joint,zhang2021joint}. In
particular, in such systems, the distributed sensors first linearly
compress their local received signals and then send them to a fusion
center for joint estimation. Considering the advantages of the LC scheme in reducing fronthaul costs and its low computational complexity, the LC-based approach is seen as a promising solution to address the limitations of fronthaul bandwidth and the computational capabilities of baseband units (BBUs) in practical ELAA systems. In light of this, the LC-based MUE (LC-MUE) schemes have been proposed in \cite{zhao2023decentralized,xu2024joint}. In what follows, we give an introduction to the LC-MUE schemes in the literature, primarily
focusing on the practical multi-carrier case \cite{xu2024joint}.

\subsubsection{LC-MUE Scheme for Multi-Carrier Systems} 
\begin{figure}[t]
	\centering
	\includegraphics[width=0.86\linewidth]{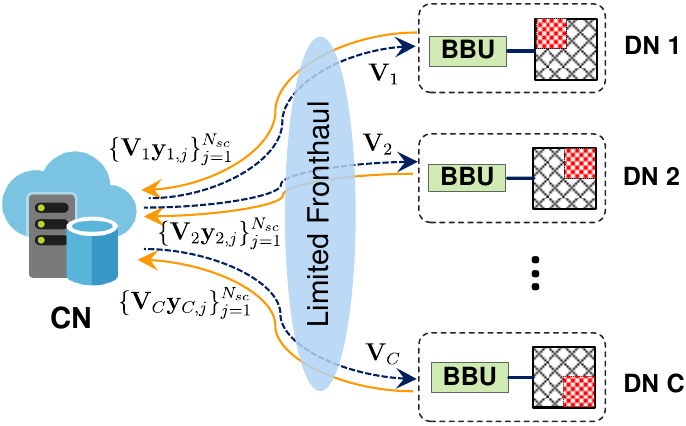}
        \captionsetup{justification=justified, singlelinecheck=false, font=small}
	\caption{\small Illustrations of the DBP architecture.} 
	\label{fig: lc-mue}
\end{figure}

The idea of the LC-MUE scheme is analogous to the LC-based distributed
parameter compression-estimation in
\cite{song2005sensors,schizas2007distributed,zhang2019joint,zhang2021joint},
which is implemented through the following steps:
\begin{itemize}
    \item[{i)}] The CN jointly designs the compressors and the equalizers for each subcarrier by solving the JCDE problem;
    \item[{ii)}] The CN sends the compressors to the DNs for compressing the received signals;
    \item[{iii)}] The DNs send the locally compressed signals to the CN for centralized equalization using the equalizers designed in step i).
\end{itemize}   
An illustration of the information exchange process of the LC-MUE
scheme is depicted in Fig. \ref{fig: lc-mue}. It is noted that, since
the number of compressors is proportional to the number of
subcarriers, the compressor may be shared across multiple subcarriers
to save the fronthaul cost.

As shown in Fig. \ref{fig: lc-mue}, the fronthaul cost of the LC-MUE
scheme consists of two parts. One is for delivering optimized
compressors from the CN to DNs, and the other is for sending the
compressed signals from DNs to the CN.  So, the total fronthaul cost
of the LC-MUE scheme is $\sum_{i=1}^C M_i r_i +
N_{sc}N_{sym}\sum_{i=1}^C r_i$ where $N_{sym}$ denotes the number of
symbols that the channel keeps invariant. Therefore, compared to the
traditional LMMSE scheme, which has a fronthaul cost of
$MN_{sc}N_{sym}$, the overall fronthaul cost of the LC-MUE scheme is
expected to be reduced substantially when $r_i$s' are small, and when
$M$ and $N_{sc}$ are large. For example, for the case with $M = 256$,
$N_{sc} = 128$, $N_{sym} = 14$, $L = 32$, $r_i = 16, \forall i$, and
$C = 4$, the fronthaul cost of the LC-MUE scheme is only $25.9\%$ of
the conventional centralized LMMSE scheme.

Based on the above model and the LMMSE criteria, the LC-MUE scheme
involves solving a multi-carrier joint compression and data
equalization (MC-JCDE) design problem to minimize the equalization
MSE, which is given by
\begin{align}\label{eqn: mse minimization problem}
	\min\limits_{\substack{\{\Ub_j\}_{j=1}^{N_{sc}},\\\Vb ~\textrm{block~diagonal}}} ~& \sum_{j=1}^{N_{sc}} \mathbb{E}\left[\left\|\sb_j-\Ub_j \Vb \yb_j\right\|^2\right], 
\end{align}
where $\Ub_j \in \Cs^{L \times r}$ is the LMMSE equalizer for each
subcarrier $j$, and $\Vb = \operatorname{blkdiag}(\Vb_1, \ldots,
\Vb_C) \in \Cs^{r \times M}$ is the compressor, which is
block-diagonal with $\Vb_i$ on the $i$-th diagonal block, $r =
\sum_{i=1}^{C} r_i$ with $r_i < M_i$ denoting the compressed dimension
of DN $i$.  It is worthy to note that when the flat-fading channel,
e.g., $N_{sc}=1$, is considered, problem \eqref{eqn: mse minimization
  problem} reduces to the problem studied in
\cite{song2005sensors,schizas2007distributed,zhang2019joint,zhang2021joint,zhao2023decentralized},
which can be efficiently solved by the block coordinate descent (BCD)
algorithm. However, the BCD algorithm is no longer efficient for
\eqref{eqn: mse minimization problem}, since updating the compressor
$\{\Vb_i\}$ involves computing a matrix inversion of dimension $M_ir_i
\times M_i r_i$. To address this issue, several low-complexity
algorithms have been proposed in \cite{xu2024joint}. In the following,
we briefly discuss these algorithms.
{\bf (1) \bf BCD-ADMM Algorithm:} 
Rather than solving problem \eqref{eqn: mse minimization problem} directly, the BCD-ADMM algorithm considers the following problem to update compressor $\Vb_i$:
\begin{subequations}\label{eqn: mse minimization trace consensus}
    \begin{align} 
        \min_{\Vb_{i,a},\Vb_{i,b}} ~&\sum\limits_{j=1}^{N_{sc}} \Tr\Big( \Ub_{i,j}^k\Vb_{i,a}\Rb_{\yb_{i,j}}\Vb_{i,b}^\Hf(\Ub_{i,j}^k)^\Hf  \notag\\
        &\qquad\quad-\mathcal{R}\big( \Ub_{i,j}^k(\Vb_{i,a} + \Vb_{i,b})\Rb^k_{\yb_{i,j}\ab_{i,j}} \big)  \Big)  \label{eqn: mse minimization trace consensus objective}\\
        \st ~&\Vb_{i,a} = \Vb_{i,b}.
    \end{align}
\end{subequations}
Problem \eqref{eqn: mse minimization trace consensus} can be solved by the nonconvex ADMM method \cite{hong2016convergence,chang2014multi}. Interestingly, as shown by \cite[Theorem 1]{xu2024joint}, the ADMM algorithm can actually converge to same $\Vb_i$ as the BCD algorithm.
Another merit of the BCD-ADMM algorithm is that it only involves low-dimensional matrix multiplications (i.e., $M_i \times M_i$ and $M_i \times r_i$), thereby having a much lower computational complexity compared to the BCD algorithm, especially when the numbers of antennas and subcarriers are large.

{\bf (2) \bf SVD-AGG Algorithm:} The SVD-AGG algorithm consists of three steps:
\begin{enumerate}
    \item[i)] A small number of subcarriers are uniformly selected from the $N_{sc}$ subcarriers, and for each subcarrier, one solves the corresponding single-carrier JCDE problem to obtain a compressor;
    \item[ii)] For each DN $i$, the obtained compressors for the selected subcarriers are aggregated by a singular value decomposition (SVD)-based scheme to produce an approximated compressor, which will be shared by all subcarriers; 
    \item[iii)] With the approximated compressor, the equalizer for each subcarrier is obtained accordingly.
\end{enumerate}
Note that the SVD-AGG algorithm first adopts the simple carrier-wise JCDE solution, followed by a succinct aggregation step to generate a high-quality shared compressor. Therefore, it enjoys further reduced complexity compared to the BCD-ADMM algorithm.

{\bf (3) \bf Fully Decentralized Scheme:} The above algorithms requires the global CSI to be available at the CN, which may cause large delay by sending the local pilot signals from DNs to the CN. The fully decentralized (FD) scheme allows the DNs relies on the local information to design a local compressor by using the previous algorithms, e.g., the BCD-ADMM algorithm. Then, the equalizer for each subcarrier is obtained accordingly by assuming a block-diagonal $\Rb_{\yb_j}$. Such an FD scheme can be found useful to initialize the previous algorithms for further computational complexity reductions.

{\bf (4) Partially Decentralized Scheme:} The partially decentralized (PD) scheme can realize more flexible tradeoff between the equalization accuracy and computational complexity. Specifically, in the PD scheme, each DN first designs a local compressor relying its local information, and sends the compressed pilot signals, together with data signals, to the DU. Then, the CN utilizes the compressed pilot signal to estimate the ``effective'' $\Rb_{\yb_j}$. Finally, the equalizer for each subcarrier is obtained by using the estimated $\Rb_{\yb_j}$ and the received compressed signals. It is noted that, if the local compressed dimension is set to the number of users and the flat-fading channel is considered, the PD scheme reduces to the ``cDR-MMSE'' scheme in \cite{zhao2023decentralized}.

\subsubsection{Future Work}
In practical TDD systems, uplink interference is estimated using the
demodulation reference signal (DMRS), which is transmitted alongside
the uplink user data. Previous schemes assume that full interference
information is available at the CN, implying that the DNs must send
the DMRSs to the CN for centralized interference estimation. Given
that the DMRS and uplink user data are transmitted together, sending
the DMRSs to the CN could incur a high processing delay, making it
unsuitable for applications with strict delay requirements. Therefore,
it is of interest to investigate efficient LC-MUE schemes that assume
that the interference signal is only available at the DNs.

\subsection{Downlink Precoding}
Downlink precoding is a fundamental SP technique in ELAA systems to
manage inter-user interference and enhance signal strength at the
receiver end, thereby increasing the system spectral efficiency. Under
the DBP architecture, the received signal at user $k$ on the $j$-th
subcarrier is given by
\begin{align} 
    \yb_k^{(j)} &\!= \Hb_k^{(j)} \Zb_k^{(j)} \sb_k^{(j)} \!+\! \sum_{\ell \neq k}^K \Hb_k^{(j)} \Zb_{\ell}^{(j)} \sb_{\ell}^{(j)} \!+\! \nb_k^{(j)} \notag\\
        & \!= \sum_{i=1}^C \Hb_{i,k}^{(j)} \Zb_{i,k}^{(j)} \sb_k^{(j)} + \sum_{\ell \neq k}^K \sum_{i=1}^C \Hb_{i,k}^{(j)} \Zb_{i,\ell}^{(j)} \sb_\ell^{(j)} \!+ \!\nb_k^{(j)},
        \label{eq:dpelaa}
\end{align}
where $\Hb_{i,k}^{(j)} \in \Cs^{N_k \times M_i}$ is the channel
between the DN $i$ and user $k$ on subcarrier $j$, $\Zb_{i,k}^{(j)}
\in \Cs^{M_i \times L_k}$ is the corresponding precoder for user $k$,
$\sb_k^{(j)} \in \Cs^{L_k \times 1}$ represents the signal for user
$k$, and $\nb_k^{(j)} \in \Cs^{N_k \times 1}$ is the AWGN at user $k$
with $N_k$ denoting the number of received antenna at user $k$. Then,
the sum-rate maximization problem can be formulated as
\begin{subequations} \label{eqn: conventioanl precoding single bs}
    \begin{align} 
		\max_{\{\Zb_{i,k}^{(j)}\}} ~& \sum_{j=1}^{N_{sc}} \sum_{k=1}^{K} R_k^{(j)}\\ 
		\st ~& \sum_{j=1}^{N_{sc}}\sum_{k=1}^{K} \sum_{i=1}^C \Tr \big(\Zb_{i,k}^{(j)} (\Zb_{i,k}^{(j)})^\Hf \big) \leq P_{\max},  \label{eq:11b}
    \end{align} 
\end{subequations}
where $P_{\max}$ is the maximum transmission power of the BS, and  
\begin{eqnarray}
    \!\!\!\!R_{k}^{(j)} \!\!\!\!&=\!\!\!\!& \log \det \big \{ \Ib + \sum_{i=1}^C \Hb_{i,k}^{(j)}\Zb_{i,k}^{(j)}(\Hb_{i,k}^{(j)}\Zb_{i,k}^{(j)} )^\Hf \cdot \notag\\
    & & \big[\sum^{K}_{\ell \neq k} \sum_{i=1}^C \Hb_{i,k}^{(j)} \Zb_{i,\ell}^{(j)} (\Hb_{i,k}^{(j)} \Zb_{i,\ell}^{(j)} )^\Hf + (\sigma_{k}^{(j)})^2 \Ib\big]^{-1}\big \}
\end{eqnarray}
is the achievable data rate of user $k$ on subcarrier $j$. 

There have been numerous studies on distributed/decentralized
precoding algorithm design based on a model similar to that of
\eqref{eqn: conventioanl precoding single bs}
\cite{muris2019fully,rusek2021tradeoff,li2018feedforward,rusek2020decentralized,li2019design,sanchez2019decentralized,zhao2023communication}. Specifically,
\cite{li2017decentralized,li2018feedforward} proposed several
distributed precoding algorithms based on Zero-Forcing (ZF) and Wiener
Filter (WF) methods, while \cite{li2019design} explored the trade-off
between computational complexity and fronthaul cost under the DBP
architecture. Recently, \cite{zhao2023communication} introduced two
decentralized precoding algorithms based on Eigen Zero-Forcing (EZF)
and weighted minimum mean square error (WMMSE) precoding schemes,
respectively, to achieve communication-efficient designs. Despite
these promising advancements, existing works primarily focus on
designing distributed algorithms for the DBP architecture without
adequately addressing the limitations of computational complexity and
fronthaul cost. Moreover, these algorithms are mainly developed for
single subcarrier scenarios and are not applicable to multi-carrier
systems with limited interconnection bandwidth, restricting their
practical applicability.

\begin{figure}[t]
	\centering
	\includegraphics[width=0.78\linewidth]{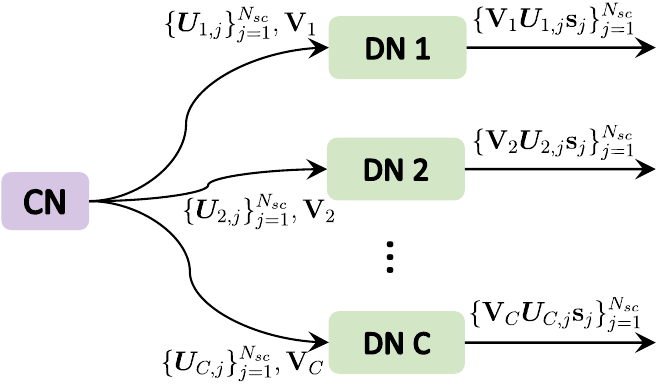}
        \captionsetup{justification=justified, singlelinecheck=false, font=small}
	\caption{\small Illustrations of the DBP architecture.} 
	\label{fig: lcp}
\end{figure}

On the other hand, it is noted that in ELAA systems, the number of
antennas in each DN can be significantly larger than the number of
data streams. This provides an opportunity to design a
compression-based precoding scheme to reduce the interconnection
cost. In the literature, compression-based methods have been studied
for distributed wireless networks to reduce fronthaul costs
\cite{zhou2014optimized,park2014fronthaul,zhou2016fronthaul,zhou2016optimal,park2013joint}. For
example, quantization-based compression methods were investigated in
\cite{zhou2014optimized,zhou2016fronthaul,zhou2016optimal} to reduce
the fronthaul cost of C-RAN with finite fronthaul capacity.  As
discussed in the subsection on uplink equalization, the LC-based
method is efficient for reducing fronthaul costs, which motivates the
LC-based precoding (LCP) scheme. The idea of the LCP scheme was first
considered in \cite{wiffen2021distributed} for a C-RAN
network. Specifically, \cite{wiffen2021distributed} addressed a joint
compressor and equalizer design problem and then directly reused the
uplink compressor for the downlink by exploiting the channel
reciprocity of TDD systems. However, this work focused only on
single-carrier systems and considered heuristics and simple ZF
precoder. Moreover, directly applying the uplink compressor to the
downlink could lead to performance loss, as the colored noise in the
downlink generally differs from that in the uplink. To overcome these
disadvantages, it is urgent to develop a more sophisticated LCP scheme
for the ELAA system using DBP architecture. Recently,
\cite{xu2024joint} considered the LCP scheme for a practical
multi-carrier ELAA system using DBP architecture, as shown in
Fig. \ref{fig: lcp}.

\subsubsection{LCP Scheme for Multi-Carrier Systems}
Similar to the uplink LC-MUE scheme, the LCP scheme for the downlink system also requires only a single roundtrip of information exchange between the CN and the DNs. The detailed workflow is outlined as follows:
\begin{enumerate}
    \item[i)] The CN jointly designs the compressor and precoder by solving a joint compression and precoding design (JCPD) problem;
    \item[ii)] The CN sends the precoder and compressed signals to the corresponding DNs;
    \item[iii)] The DNs precode the compressed signals with the precoder and send the precoded signal to the users.
\end{enumerate}
In this scheme, the signal on each subcarrier will be designed a
compressor, while the precoder may be shared by multiple subcarriers
for fronthaul reduction, since the precoder should be sent to the
DNs. It is noted that the key of the LCP scheme is to efficiently
solve the JCPD problem, which is given by
\begin{subequations} \label{eqn: LCP single bs}
    \begin{align}
        \max _{\substack{\{\Ub^j\}_{j=1}^{N_{sc}},\\\Vb ~\textrm{block~diagonal}}}~& \sum^{N_{sc}}_{j=1} \sum^{K}_{k=1} R_{k}^{j}\\
        \st ~& \sum^{N_{sc}}_{j=1} \sum_{k=1}^{K} \sum_{i=1}^C \|\Vb_i^\Hf \Ub_{i,k}^{(j)}\|_2^2 \leq P_{\max}, \label{eq:12b}
    \end{align}
\end{subequations}
where {\small $R_{k}^{j}$} is the achievable rate of user $k$ on
subcarrier $j$, which is given by {\small $R_{k}^{(j)} = \log \det
  \big \{ \Ib + \sum_{i=1}^C \Hb_{i,k}^{(j)} \Vb_i \Ub_{i,k}^{(j)} $}
{\small$(\Hb_{i,k}^{(j)} \Vb_i \Ub_{i,k}^{(j)} )^\Hf
  \big[\sum^{K}_{\ell \neq k} \sum_{i=1}^C \Hb_{i,k}^{(j)} \Vb_i
    \Ub_{i,\ell}^{(j)} (\Hb_{i,k}^{(j)} \Vb_i \Ub_{i,\ell}^{(j)} )^\Hf
    + (\sigma_{k}^{(j)})^2 \Ib\big]^{-1}\big \} $}, $\Ub_{i,k}^j \in
\Cs^{L \times r_i}$ is the compressor for user $k$ on subcarrier $j$,
and $\Vb_i \in \Cs^{r_i \times M}$ is the shared precoder.

Compared to \eqref{eqn: conventioanl precoding single bs}, the LCP
problem \eqref{eqn: LCP single bs} is much more challenging to solve
owing to the coupling of the compressors and precoder in both the
objective function and constraints \cite{liu2024survey}.  To
solve problem \eqref{eqn: LCP single bs} efficiently,
\cite{shi2024linear} proposed two algorithms based on a penalty dual
decomposition (PDD) method and a matrix factorization (MF) method,
respectively. Specifically, the PDD-based algorithm jointly optimizes
the compressor and precoder, which achieves a promising performance
but with a high computational complexity. While the MF-based algorithm
aims to solve problem \eqref{eqn: LCP single bs} in a low-complexity
manner. In what follows, we give a brief introduction to the MF-based
algorithm

Basically, the MF-based algorithm consists of the following two steps:
\begin{itemize}
    \item[i)] One solves the conventional precoding design problem in \eqref{eqn: conventioanl precoding single bs} to obtain a set of precoders, $\Zb_k^{(j)}$,  for all subcarriers. In particular, the precoders can be obtained by using the well-known WMMSE algorithm \cite{shi2011iteratively} or the EZF algorithm \cite{sun2010eigen}.
    \item[ii)] With the attained precoders, one solves the following low-rank MF problem to obtain the compressors and the shared precoder
    \begin{align} \label{eqn: MF}
        \min_{\{\Vb_i, \widetilde\Ub_i \}}  ~
        \sum_{i=1}^C \|\widetilde\Zb_i - \Vb_i \widetilde\Ub_i\|_F^2, 
    \end{align} 
    where 
    {\small$\widetilde \Zb_i = [\Zb_i^{(1)}, \ldots,\Zb_i^{(N_{sc})}] \in \mathbb{C}^{M_i \times LN_{sc}}$}, 
    and {\small$\widetilde\Ub_i = [\Ub_i^{(1)}, \ldots,\Ub_{i}^{(N_{sc})}] \in \mathbb{C}^{r_i \times LN_{sc}}$}.
    To obtain the factorization results efficiently, the SVD operation is performed on $\widetilde\Zb_i$ and utilize the first $r_i$ columns of right singular matrix to construct $\widetilde \Ub_i$. 
    The corresponding $\Vb_i$ is determined by the least squares solution.
    Finally, the $\Ub_{i,k}^{(j)}$ are scaled to ensure that the power constraints are active.
\end{itemize}
The MF-based algorithm has comparable computational complexity to the
conventional centralized precoding scheme, since the second step only
performs a low-complexity MF operation, making it preferable in
practical applications where powerful servers are not available.

\subsection{Accuracy of the CSI, and Phase Calibration}\label{sec:synccalibr}

There are four different error sources in the CSI (channel
estimates) in large-scale and distributed MIMO: (a) noise and
interference on the channel estimates; (b) channel aging (the fact
that estimates become outdated); (c) errors resulting from
uplink-downlink reciprocity imbalances within an array; and (d) errors
resulting from oscillator drifts (accumulated phase noise) when
different service antennas in an array, or different distributed
arrays, are driven by oscillators that are not mutually
phase-locked. In what follows, we discuss these different errors
sources and how they can be mitigated:
\begin{enumerate}
\item[(a) \bf{Noisy channel estimates:}] Noisy channel
  estimates can be a significantly limiting factor, especially on
  uplink where the available power is generally smaller. The
  countermeasures are to increase power (to the extent permitted) and
  to increase the integration gain (length of the pilot sequences);
  see \cite{fundamentalsofmassivemimo}.

Interference on the channel estimates can originate either from reuse
of pilots or from data transmission on the pilot dimensions in other
cells; both are equally bad
\cite[Sec.~4.4.3]{fundamentalsofmassivemimo}. The countermeasures are
sparse enough pilot reuse \cite[Chap.~6]{fundamentalsofmassivemimo},
and the use of second-order statistics of the fading process to
improve the channel estimates \cite{bjornson2017massive} --
effectively, spatially filtering the received pilots before estimating
the channels.

\item[(b) \bf{Channel aging:}] In most analyses of massive and
  distributed MIMO, the channel is modeled as block-fading.  While
  this is an excellent abstraction for capacity analysis and system
  optimization tasks such as power control, in reality the channel
  response varies continuously as function of time and frequency.  On
  uplink, given pilots that are spread out over the time-frequency
  domain, one can easily interpolate both over time and frequency to
  obtain estimates for each time-frequency resource element
  (sample). The downlink [in TDD] is more difficult, as one has to
  rely on prediction, or on the (erroneous) presumption that the
  channel remains static once transmission has switched from uplink to
  downlink. Channel prediction is an active area of research; for some
  early contributions see \cite{wu2019inverse}.

\item[(c) \bf{Uplink-downlink reciprocity imbalance:}] Within
  an array, where all antennas are driven by the same local
  oscillator, there may be an imbalance between the phase lag in the
  receive and in the transmit branch; this imbalance nominally breaks
  the uplink-downlink reciprocity required for reciprocity-based
  multiuser MIMO beamforming in TDD to work.  The countermeasure is
  reciprocity calibration. State-of-the-art methods rely on
  transmission between different antennas \emph{within} the array, and
  taking bidirectional measurements from which the reciprocity
  imbalances can be estimated
  \cite{kaltenberger2010relative,vieira2017reciprocity,zetterberg2011experimental,chen2022hierarchical,shepard2012argos,lee2017calibration,JiangKDLK2015,luo2019massive,papadopoulos2014avalanche,jiang2018framework}.
  This calibration can done rather infrequently, since the
  uplink-downlink phase imbalances only change slowly over time (they
  are mainly dependent on external factors such as temperature).

\item[(d) \bf{Oscillator drifts:}] When different antennas in
  an array, or different panels (subarrays) are driven by independent
  local oscillators that are not locked in phase, calibration
  measurements are required to \emph{jointly reciprocity-calibrating}
  the arrays.\footnote{Also known as phase synchronization, or phase
  alignment.} This can be done by bidirectional over-the-air
  measurements \emph{between the arrays}
  \cite{rogalin2014scalable,vieralarsson_pimrc,chen2017distributed,kim2022gradual,cao2023experimental,balan2013airsync,rashid2022frequency,ganesan2023beamsyncTWC}.
  This calibration has to be re-done every time the oscillators have
  drifted by some amount, which in practice means very frequently
  unless very stable (and expensive) oscillators are used: In contrast
  to the reciprocity-calibration within an array [(c) above], this
  calibration may have to be performed on a millisecond-level
  timescale.  An estimation-theoretic analysis of over-the-air
  calibration of distributed MIMO, based on a graph representation of
  the who-measures-on-whom topology, can be found in
  \cite{larsson2024massive}. Therein it is shown, in particular -- and
  perhaps counterintuitively -- that the more antennas that are
  involved in the calibration the less accurate the phase estimates
  become, but the more accurate the beamforming becomes.

\end{enumerate}
Importantly, (a)--(d) affect the downlink, but only (a)--(b) affect
the uplink since on uplink, pilots and data see the same channel, so
neither reciprocity nor phase alignment between antennas is required.
Especially the fourth effect, (d), appears to have been the source of
significant misconceptions in the literature, as discussed in
\cite{nissel2022correctly,larsson2023phase}.

It should be noted in this context that synchronization in an ELAA system entails several different components:
synchronization in frequency between independent panels; synchronization in time; synchronization of the sampling clocks;
and synchronization (alignment) of the phase for joint coherent downlink beamforming. Among these problems, it is
the phase synchronization, discussed above under item d), that is the most difficult. However, open problems remain especially
on how to efficiently synchronize the sampling clocks.

Given the importance of synchronization, especially phase alignment among distributed panels, the development
of robust and efficient methods should be a priority for future work. For example,
when introducing phase alignment measurements into the picture, the
TDD flow must be broken.  Specifically, an access point that is
receiving calibration signals from another access point needs to
reverse its operation in the sense that uplink and downlink are
swapped: during one or multiple slots it must listen when other access
points transmit and vice versa. This creates a host of new problems.
One is that an access point that switches from transmission to reception during (a
part of) a slot nominally assigned to downlink transmission will miss
the opportunity to transmit on downlink to users during that part of
the slot. This will affect the quality of service of the traffic,
especially for low-latency applications and for cases when the access
point in question is critical to or the \emph{only one} of the access
points serving a particular user.
An optimization framework could be devised to determine which access
points should transmit calibration signals and which access points
should receive calibration signals in a given slot, taking into account the requirements of the users being
served at a given point in time, the traffic situation, and the
traffic's latency constraints.

\section{Distributed SP for ELAA Systems with Coordinated Distributed Antennas} \label{sec: distributed antenna}
In this section, we focus on the distributed SP algorithm designs for
ELAA systems with coordinated distributed antennas. Specifically, we
concentrate on two representative systems: the multi-cell coordinated
transmission system and the CF-mMIMO system. In multi-cell coordinated
transmission systems, multiple massive MIMO BSs coordinate to
jointly serve multiple users in these cells. In CF-mMIMO systems, a
central processing unit coordinates multiple APs to jointly serve
multiple users in a spatial area. Since the CF-mMIMO system performs
user-centric transmission, it can provide uniform performance across
the network.

\subsection{Similarities and Differences With the Single-BS ELAA Systems Using DBP Architectures}
The considered ELAA systems with distributed antennas share a similar
system architecture with single-BS ELAA systems using DBP architectures.
Both system forms leverage the DBP architecture, allowing for
distributed signal processing (SP) across multiple processing
units. Consequently, many algorithms developed for single-BS ELAA systems
can be applied to systems with distributed antennas. However, there
are still two key differences that necessitate further investigations.
\begin{itemize}
    \item Firstly, in single-BS ELAA systems, it is assumed that there is
      always a central server (i.e., CN) capable of performing
      centralized processing. However, such a powerful central server
      may not always exist in coordinated multi-cell transmission
      systems, necessitating the development of more efficient
      distributed SP algorithm designs
      \cite{Tolli2011decentralized,antonioli2020decentralized,cai2023approaching}.
    \item Secondly, while antennas in single-BS ELAA systems are
      co-located, it is no longer the case in ELAA systems with
      distributed antennas. Under distributed antenna settings, the
      channels between users and some antennas may be poor due to
      large-scale fading. Consequently, it is not efficient to use all
      antennas to serve all users, particularly when considering
      computation- and communication-efficient distributed SP designs.
      Therefore, this spatial distribution introduces new research
      challenges, such as user association and user scheduling.
\end{itemize}
These two differences pose new research challenges, leading to
distinct research problems that require innovative
solutions. Addressing these challenges will be the focus of the
subsequent subsections, where we will explore the developments of
efficient distributed SP algorithms tailored to ELAA systems with
coordinated distributed antennas.


\subsection{Distributed SP for Multi-Cell Coordinated Transmission Systems}
In multi-cell coordinated transmission systems, the distributed
downlink precoding problem has been widely studied in the
literature. However, the distributed channel estimation and uplink
equalization problems have been rarely addressed. In this subsection,
we primarily focus on an overview of the distributed downlink
precoding algorithms and distributed user scheduling algorithms for
multi-cell coordinated transmission system.

\subsubsection{Review of Existing Distributed Multi-Cell Coordinated Precoding Algorithms}
Consider a downlink multi-cell system where $C$ BSs cooperative to
serve multiple users. For ease of illustration, we assume that one
user is scheduled in each cell for downlink transmission. Then, the
centralized multi-cell coordinated precoding (MCP) problem in
flat-fading channel is given by
\begin{subequations} \label{eqn: conventioanl precoding multi cell}
    \begin{align} 
		\max_{\{\Zb_{i}\}}~ & \sum_{i=1}^{C} R_{i}\\ 
		\st ~&  \Tr \big(\Zb_i \Zb_i^\Hf \big) \leq P_{\max}, \forall i \in \mathcal{C},
    \end{align}
\end{subequations}
where $\Zb_i$ is the precoder for the user in cell $i$, and
\begin{eqnarray}
R_{i} & = & \log \det \big \{ \Ib + \Hb_{i,i}\Zb_i (\Hb_{i,i}\Zb_i)^\Hf \cdot \nonumber \\ & & \big[\sum_{\ell \neq i}^{C} \Hb_{i,\ell} \Zb_{\ell} (\Hb_{i,\ell} \Zb_{\ell})^\Hf + \sigma_i^2 \Ib\big]^{-1}\big \}  \label{eq:ardw}
\end{eqnarray}
is the achievable data rate of cell $i$. Here, $\Hb_{i,\ell}$ is the
channel between BS $\ell$ and the user in cell $i$.

The MCP problem has been widely studied in the literature
\cite{gesbert2010multi,shi2011iteratively,hong2013joint,park2013joint,nigam2014coordinated,park2015cooperative},
which showed that MCP is able to eliminate both intra-cell and
inter-cell interference, thereby providing considerable performance
gains. However, as the number of antennas and users increases, these
cooperative strategies may become less desirable due to the overheads
associated with channel information exchange among BSs via backhaul
links and the need for a powerful central processor. Additionally, the
computational complexities of the algorithms scale rapidly with the
network size. A naive approach to overcoming these challenges is to
perform fully decentralized precoding, where each BS optimizes the
precoders for the users in its cell solely based local acquired CSI,
while treating inter-cell interference as noise. However, in practical
scenarios, the inter-cell interference is actually a bottleneck of
multi-cell system, and thus the performance of fully decentralized
schemes could be significantly degraded. Therefore, efficient
distributed precoding algorithms, which can not only achieve promising
performance but also have low inter-connection costs, are crucial for
making multi-cell coordinated transmission practical.

In the literature, there have been many works studying efficient
distributed MCP (D-MCP) designs
\cite{shen2012distributed,Maros2018admm,boukhedimi2017coordinated,li2022decentralized,han2020distributed,pennanen2011decentralized,Tolli2011decentralized},
and several promising approaches have proposed. In what follows, we
give an introduction to three efficient distributed precoding
approaches aiming at reducing the cost of backhaul signaling:
\begin{itemize}
    \item {\bf D-MCP via new metrics:} When the system cannot capture
      interference from other cells, it is possible to improve system
      performance by reducing leakage to other cells only based on
      local CSI. Motivated by this, the work
      \cite{boukhedimi2017coordinated} proposed a new metric, called
      signal-to-leakage-and-noise ratio (SLNR), which involves the
      amount of interference caused by desired signal on other
      users. Different from the conventional SINR-based schemes, the
      SLNR metric can decouple the optimization problem and allows
      fully decentralized solutions. The concept of leakage was
      further developed in \cite{li2022decentralized} where the
      signal-to-leakage-plus-interference-plus-noise ratio (SLINR) is
      defined by simultaneously considering both the intra-cell
      interference and the total interference leakage to users served
      by other cells. The work \cite{han2020distributed}
      defined another expression of SLINR by changing the sum of
      interference leakages to the geometric mean. The associated
      problems can be efficiently solved by the WMMSE method. However,
      the discrepancy between the SLINR and the SINR may lead to
      performance loss in practice. Besides, the geometric-mean
      leakage term in \cite{han2020distributed} is expensive to
      compute and causes large gradient-computation overhead.
    
    \item {\bf D-MCP via user cooperation:} Considering a TDD systems,
      the work \cite{atzeni2020distributed} proposed a distributed
      precoding scheme with a new information exchange mechanism via
      user cooperation. In particular, the proposed scheme introduced
      a new uplink signaling resource and a new CSI combining
      mechanism that complement the existing uplink and downlink
      pilot-aided channel estimations. This enables each BS to acquire
      the necessary cross-term information, entirely eliminating the
      need for backhaul signaling for CSI exchange. However, the
      proposed algorithm in \cite{atzeni2020distributed} is iterative
      in nature, which may cause a large processing delay.
    
    \item {\bf D-MCP via interference power exchange:} Another
      approach to reduce the backhaul signaling is to exchange the
      interference power (IFP) instead of CSI between the cooperative
      BSs
      \cite{pennanen2011decentralized,Tolli2011decentralized}. This
      leads to a problem that jointly optimize the precoding and IFP
      variables, which can be iteratively solved by the primal
      decomposition method \cite{pennanen2011decentralized} or the
      dual decomposition method \cite{Tolli2011decentralized}. During
      the iterations, BSs that interfere with each other only need to
      the exchange IFP variables. However, in practical system, the
      IFP graph maynot be the same as the backhaul graph, which brings
      new challenges in IFP variables exchange among BSs.
\end{itemize}
Recently, a simple yet efficient D-MCP scheme, relying on a virtual
power control (VPC) based formulation, was proposed in
\cite{cai2023approaching}. The VPC-based formulation is similar to the
centralized formulation but is solved by each BS individually based on
its own CSI. Additionally, the cooperative BSs only need to exchange
the interference channels to users in other cells and transmission
powers. As a result, the performance can approach that of the
centralized scheme but with a much reduced backhaul cost. In what
follows, we give a more detailed introduction to the VPC-based D-MCP
scheme.

\subsubsection{Efficient D-MCP via a VPC-based Formulation} 
As seen from \eqref{eqn: conventioanl precoding multi cell}, once the
interference terms, i.e., $\Hb_{i,\ell} \Zb_{\ell}, \forall \ell \neq
i$, are given, each BS can solve the problem by itself based on local
CSI.  On the other hand, the interference terms can be split into two
parts as $\Hb_{i,\ell} \Zb_{\ell} = \Gb_{i,\ell} \Pb_\ell$. Here,
$\Gb_{i,\ell}$ and $\Pb_\ell$ are respectively given by
\begin{subequations}
   \begin{align}
        \Gb_{i,\ell} &= \Big[\Hb_{i,\ell}\frac{[\Zb_{\ell}]_1}{\|[\Zb_{\ell}]_1\|},...,\Hb_{i,\ell}\frac{[\Zb_{\ell}]_M}{\|[\Zb_{\ell}]_M\|}\Big], \\
        \Pb_\ell &= \blkdiag\{\|[\Zb_{\ell}]_1\|,...,\|[\Zb_{\ell}]_M\|\},
    \end{align} 
\end{subequations}
As seen, $\Gb_{i,\ell}$ collects the corresponding interference
channels from BS $\ell$ to the user in cell $i$, and $\Pb_\ell$
represents the transmission power to the user in cell $\ell$. The idea
of the VPC-based scheme is to allow each BS to ``virtually'' control
the transmission power variables of other BSs so as to predict their
behaviors as if the centralized optimization is conducted. In
particular, by inserting $\Gb_{i,\ell}$ and $\Pb_\ell$ into $R_{i}$ in
problem \eqref{eqn: conventioanl precoding multi cell}, the
central-like VPC-based sum rate maximization problem to be solved by
BS $i$ can be written as
\begin{subequations} \label{eqn: vpc precoding multi cell}
    \begin{align} 
		\max_{\Zb_{i},\Pb_{\ell}} ~& R_i(\Zb_i,\Pb_\ell)\\ 
		\st ~&  \Tr \big(\Zb_i \Zb_i^\Hf \big) \leq P_{\max},\\
                ~& \Tr \big(\Pb_\ell \Pb_\ell^\Hf \big) \leq P_{\max}, \forall \ell \neq i,
    \end{align}
\end{subequations}
which can be solved by the WMMSE method \cite{cai2023approaching}. 

It is noted that, by the VPC-based scheme, only the interference
channels and the transmission powers are required to be exchanged
between the BSs. Since these two terms are not related to the number
of transmit antennas, the backhaul cost of the VPC-based scheme is
relatively low, especially for the ELAA systems where the transmit
antenna is large.

\subsubsection{Sensing-Assisted Distributed User Scheduling}
Distributed user scheduling with the merits of distributed processing
also plays an important role to improve the network spectral
efficiency while incurring a low backhaul cost
\cite{khan2018optimizing,ng2011resource,li2022decentralized}.  For
most existing distributed algorithms, including that discussed in the
previous subsection, the CSIs or inter-cell interference powers are
still required to periodically exchanged between the BSs in each
iteration of the algorithm. However, in time-varying environments,
frequent message exchanges will cause extra delays, and the BSs would
have outdated information. The SLINR-based scheme, which only needs
the intra-cell interference and the leakage to users in other cells,
is an efficient way to alleviate this issue
\cite{cai2023approaching,han2020distributed}. However, the coupling of
user scheduling variables in the SLINR term causes the difficulty of
estimating the cross-cell leakage. The work \cite{li2022decentralized}
further introduced a traffic model to estimate the leakage
approximately without requiring knowing the scheduling variables of
other cells. However, the traffic model in \cite{li2022decentralized}
cannot offer real-time leakage estimate, which can cause a discrepancy
between SLINR and the true SINR, degrading the performance in dynamic
environments.

\begin{figure}[t]
	\centering
	\includegraphics[width=0.98\linewidth]{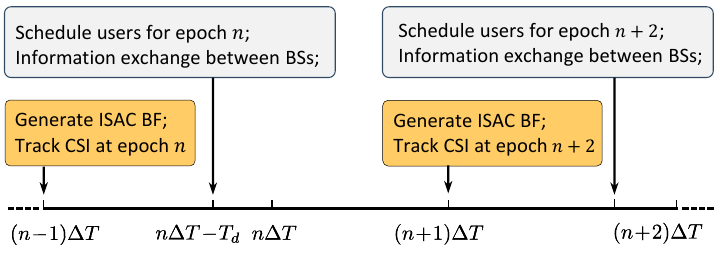}
        \captionsetup{justification=justified, singlelinecheck=false, font=small}
	\caption{\small Illustrations of the ISAC-based distributed user scheduling and beamforming design framework.} 
	\label{fig: ic-mcra}
\end{figure}

To address these challenges, \cite{cai2024sensing} proposed an
ISAC-based distributed user scheduling and beamforming design
framework that can provide close-to-centralized performance with
significantly reduced information exchange overhead. In particular,
as shown in Fig. \ref{fig: ic-mcra}, \cite{cai2024sensing} considered
a transmission time $T$, which is divided into $N$ equal-length
epochs, and the length of each is $\Delta T$. Based on this setting,
the proposed ISAC-based distributed user scheduling and beamforming
design framework consists of four steps in each epoch:
\begin{itemize}
    \item[i) \bf{ISAC Signal Transmission:}] At time $(n-1) \Delta T$,
      each BS transmits the data signal with ISAC beamformers. Based
      on the reflected signal, each BS can obtain the estimated
      kinematic parameters and the corresponding channel estimate of
      the served users;
    \item[ii) \bf{Per-cell Scheduling:}] At time $(n\Delta T - T_d)$,
      by utilizing the estimated channel, each BS conducts a user
      scheduling scheme to determine the scheduled user set for epoch
      $n$.
    \item[iii) \bf{Information Exchange:}] After computing the
      coordinates of scheduled users from the angle and distance
      information, each BS exchanges their coordinates and velocity
      information to other BSs, which will be used by other BSs to
      estimate its cross-cell LoS channel.
    \item[iv) \bf{Distributed ISAC Beamforming:}] At time $n\Delta T$,
      with the channel estimates of its served users and its
      cross-cell channels, each BS optimizes ISAC transmit beamformers
      by solving a sum rate maximization problem.
\end{itemize}
This scheme can adaptively track the users' kinematic parameters in a
dynamic environment and realize accurate channel estimation, enabling
high-quality user scheduling and beamforming designs. Besides, this
scheme also avoids the need for extensive CSI exchanges between BSs,
thereby reducing the backhaul cost.

There exist several interesting directions for future research. First, improving sensing performance in scenarios where line-of-sight (LoS) channels are unavailable is essential. Such environments, including urban canyons or indoor settings, require advanced sensing techniques that can effectively utilize non-LoS signals, for accurate user localization. Second, developing robust distributed user scheduling and beamforming algorithms that account for sensing errors is equally important. These algorithms should be capable of maintaining high network performance and mitigating the adverse effects of inaccuracies in sensing data, ensuring reliability in dynamic and challenging environments.

\subsection{Distributed SP for CF-mMIMO System}
In the considered CF-mMIMO system, a CPU coordinates $C$ multi-antenna
APs to cooperatively serve $K$ users is a certain area. The CF-mMIMO
system has a quite similar architecture to the single-BS ELAA system
using a star-networked DBP architecture, except for that the APs (can
be viewed as DNs) are spatially distributed. Then, it depends on the
functional split between CPU and APs with options ranging from 7.2 and
8 in \cite{demir2024cell} to fully distributed in \cite{Lee24}.

In functional split option 8, the processing down to the baseband
signal is performed in the CPU. The APs only perform RF processing and
transmit (receive) the pure sampled and quantized baseband signals to
(from) the CPU via fronthaul links. Therefore, option 8 has the lowest
processing power requirements of the APs. However, the rate
requirements of the fronthaul links are high. Lower fronthaul
requirements has option 7.2. Low PHY functions are implemented at the
AP. Joint processing can still be performed at the CPU while the
complexity of the APs is kept low. In contrast, the fully distributed
functional split architecture in \cite{Lee24} implements both high and
low PHY functions. One speciality is the option to control lower PHY
functions of several cluster APs by one high PHY function in one
cluster head AP.

Distributed SP is important for CF-mMIMO system for improving the
system scalability
\cite{bjorson2020making,interdonato2020local,atzeni2021distributed}. The
work \cite{bjorson2020making} investigates uplink equalization problem
in the CF-mMIMO system and proposes three distributed algorithms based
on linear equalizers, which only requires the DNs to send low-dimensional signals (e.g., the local estimates) to the CN for fronthaul reduction. 
The works \cite{interdonato2020local,atzeni2021distributed} study the
distributed downlink precoding problem in CF-mMIMO systems. The
complexity of the CF-mMIMO system scales rapidly with the network
size. To address this issue, user-centric CF-mMIMO systems with
user-AP association have been proposed, which improve the system
scalability \cite{bjornson2020scalable,d2020user}.

It depends on the constraints of fronthaul and midhaul links, whether
the signal processing, such as channel estimation, pre- and de-coding,
equalization, modulation and demodulation, coding and decoding, can be
performed in a centralized or de-centralized manner. For the uplink
transmission, the signal processing for MMSE and maximum likelihood in
the uplink can be performed decentralized without loss of optimality
\cite{Shaik2021}. There, it is shown that in the sequential processing
\cite[Algorithm 1]{Shaik2021} the estimate obtained at one AP is
equivalent to that obtained by centralized processing with LMMSE
receiver. For the downlink, the team precoding framework in
\cite{Miretti2022} allows an optimal distributed spatial pre-coding
and power control.

The main challenge in CF-mMIMO is to achieve the benefits of cell-free
operation in a practical way, with computational complexity and
fronthaul requirements that are scalable to enable massively large
networks with many mobile devices \cite{demir2021foundations}. The
monograph \cite{demir2021foundations} describes the state-of-the-art
signal processing algorithms for channel estimation, uplink data
reception, and downlink data transmission with either centralized or
distributed implementation.

Recently, the work \cite{ammar2022distributed} investigated the
distributed resource allocation for a user-centric CF-mMIMO system,
where two distributed algorithms are propose to jointly optimize the
user association and precoding. The SLNR-based metric is also applied
for backhaul signaling reduction. In \cite{Li2024}, a joint
optimization of fronthaul load and computation resources in CF-mMIMO
networks is performed. It is one the first papers to consider a graph
model for the fronthaul network jointly with the cell-free wireless
access.

The relation between CF-mMIMO and Open-RAN together with a review of
next generation multiple access methods is provided in
\cite{Jorswieck2024a}.

\subsubsection{Future Work} 

CF-mMIMO has received significant attention from the signal processing perspective. In particular, the distributed channel estimation and signal detection both for up- and downlink were studied in detail \cite{Elhoushy2022}. However, the constraints and requirements from the fronthaul network leave several open research questions for future work. 

What is the optimal choice of the functional split and corresponding distributed signal processing given fronthaul constraints and user rate requirements? Since demands and requirements are time variant, a flexible functional split would help to improve the efficiency \cite{Diez2021}. Enhanced coding schemes, such as rate-splitting can lower the fronthaul requirements and improve the achievable rates in CF-mMIMO \cite{Jorswieck2024}.

What is the impact of impairments in the fronthaul links and the RF-chains at the APs? How can we develop robust precoding and resource allocation schemes which are resilient against impairments and faults on the fronthaul and wireless links? The effects of dirty RF, such as phase noise, on the performance of CF-mMIMO is studied in recent works \cite{Fang2021}. Severe fronthaul limitations are considered in \cite{Ando2021}, and robust receivers are developed.

How can a deployment of CF-mMIMO establish near-field beamforming gains and high localization precision? In \cite{Cui2023}, the near-field MIMO communications is motivated by CF-mMIMO among others. Depending on carrier frequencies and antenna as well as AP spacing, user terminals in the near-field can enjoy highly directive and selective receive signal powers.

\section{Distributed SP for ELAA Systems Integrated with Emerging Technologies} \label{sec: emerging technology}
Recently, there has been a growing trend to integrate emerging technologies into ELAA systems to further enhance system performance. These integrations aim to improve spectral efficiency, expand wireless coverage, and enable new functionalities such as wireless sensing. In this section, we explore the interplay between ELAA systems and several key emerging technologies. Specifically, we focus on four prominent technologies: repeater-assisted multi-user MIMO, backscatter communication, RIS, and ISAC. We will delve into the advantages these integrations offer and examine the potential trade-offs and challenges they introduce.

\subsection{Repeater-Assisted Multi-User MIMO}

Cellular-massive-MIMO \cite{fundamentalsofmassivemimo}, the backbone
of the 5G physical layer, is a very powerful technology. But coverage
holes, and difficulties to send multiple streams to multi-antenna
users because of insufficient channel rank, remain issues.  The
ultimate solution will be distributed and extremely-large scale MIMO,
but while these are at heart strong technologies, installing fronthaul
and backhaul can be expensive, and achieving accurate phase-alignment
for coherent multiuser beamforming on downlink remains a difficult
technical problem (see Section~\ref{sec:synccalibr}).

Ways of improving performance by \emph{augmenting the propagation
environment} with new paths are therefore desirable.  One option is to
deploy RISs \cite{2007.03435,2203.03176} -- but they have large,
perhaps unrealistically large, form factors and require significant
amounts of training and control overhead. Also, probably, in practice,
some form of active filtering is required to make them sufficiently
band-selective.

A different, radically new approach is to deploy large numbers of
physically small and cheap wireless repeaters, that receive and
instantaneously retransmit signals -- appearing as if they were
ordinary scatterers in the channel, but with amplification
\cite{2406.00142}. Repeaters, as such, are deployed today already but
only in niche use cases (for example, in tunnels). Could they be
deployed at scale, in swarms, within the cells? What would be required
of the repeaters, and how well could a repeater-assisted cellular
massive MIMO system work, compared to distributed MIMO? What are the
fundamental limits of this technology?

Among several different possible designs, the most promising one is
dual-antenna repeaters that switch between two separate amplification
paths (one path from a first to a second antenna, and one path from
the second antenna to the first), in alignment with the TDD pattern.
Such repeaters can, with appropriate calibration \cite{10485004}, be
made \emph{reciprocal}, such that they appear \emph{transparently} to
the network as channel scatterers with amplification.  In particular,
this makes reciprocity-based multiuser MIMO beamforming
\cite{fundamentalsofmassivemimo} work, by relying on uplink channel
estimates for the downlink beamforming. This way, the repeater can do
the job of a RIS but with essentially no control overhead, and a
two-order-of-magnitude smaller form factor.

An important consideration when deploying repeater swarms within a
cell is that depending on the repeater amplification levels, the
repeaters will mutually interfere and amplify the signals transmitted
by each other.  One must ensure that the amplification levels are set
such that the positive feedback loops created between the repeaters
are stable (within an appropriate safety margin). Optimal
configuration of repeater swarms under such constraints is an open
problem; some initial analysis from a control-theoretic viewpoint can
be found in \cite{2405.01074}.

Another open problem is to optimize power allocation and receiver
algorithms to minimize the effects of amplified noise. Repeaters,
being active devices with power amplifiers, will introduce thermal
noise that gets amplified and transmitted, and received at the base
station arrays. This noise can be mitigated, and amplification levels
and activation strategies can be optimized to minimize its impact.

As demonstrated in \cite{2406.00142}, repeater-assisted cellular MIMO
could approach distributed-MIMO performance, but with much simpler
deployments that require no backhaul and no phase alignment.  There
are significant opportunities to innovate new solutions within this
nascent area of physical-layer wireless research.

\subsection{Backscattering  Communications}

Energy-neutral wireless devices, also sometimes known as ``passive''
devices, do not have a battery but rely on harvesting of
radio-frequency power for their operation. These devices communicate
through backscattering and do not have a conventional radio-frequency
frontend, but only an antenna whose load impedance can be
controlled. By controlling this impedance, the device can change the
phase of the wave scattered from its antenna when illuminated with an
incoming wave.  There are three main architectures for backscattering
communication: (a) In a {monostatic} setup, the transmitter and
receiver are co-located and share the same antenna antennas
\cite{liu2014multi,mishra2019optimal,KashyapBL2016}; such systems
require full-duplex technology at the arrays. (b) In a {bistatic}
setup, the transmitter and receiver are spatially separated and do not
share circuitry \cite{kimionis2014increased,hua2020bistatic};
full-duplex technology is not required.  (c) An {ambient}
backscattering system does not have a dedicated transmitter but relies
on ambient radio sources such as Bluetooth, WiFi, or TV broadcast
signals \cite{basharat2021reconfigurable}.
                  
The bistatic setup has the largest potential and integrates naturally
with extremely large arrays distributed MIMO, with pairs of MIMO
panels acting as transmitter and receiver.

The received backscattered signal is typically weak compared to the
direct link between the transmitter and receiver. This requires a high
dynamic range of reader circuitry \cite{biswas2021direct}, mitigation
techniques \cite{varshney2017lorea,li2019capacity} or the use of
advanced beamforming techniques to suppress direct link interference
\cite{kaplan2023direct}.  Direct link interference cancellation for
bistatic setups brings several new elements: for example, there can be
a carrier frequency or phase offset between the panels
\cite{tao2021novel}. In addition, these panels are not necessarily
synchronized in phase (cf.~Section~\ref{sec:synccalibr}).
        
The service of energy-neutral (passive) devices is challenging owing
to the poor link budget: the path gains from the transmitter to the
device and from the device to the receiver multiply. This requires
sensitive and efficient signal processing algorithms. Some specific,
new problems that emerge are:

{\bf i) Robust beamforming for initial access:} In practice, when
beamforming to a passive device, only a rough estimate of the channel
may be available.  For example, the channel could be estimated at some
point in time, but be outdated. The question then arises as to how
beamforming could be made more robust: rather than sending a single
beam obtained from the outdated channel estimate, could several beams
be formed whose combined action provide sufficient power to the
device?
 
One may, for example, employ beam diversity \cite{deutschmann}.  More
sophisticatedly, suppose we have access to an estimate of the channel
response, say $\bg$, at a nominal focal point. Suppose in addition
that we could predict how this channel response varies within some
neighborhood of that nominal focal point, and that we take samples
from this neighborhood.  Such prediction could be performed, for
example, by exploiting a geometric model. Alternatively, it could be
based on historical channel state information, by constructing a
database of past impulse responses stored in chronological order,
forming a \emph{data-driven parameterization} of a neighborhoods
around nominally measured channel responses.

In a nutshell, the problem is to find a beam $\bw$ that maximizes
$\min_i | \bg^H_i \bw |$, given channel vectors $\{\bg_i\}$ sampled
from the neighborhood. This problem is reminiscent of multicast
beamforming and can be case as a maximin matrix optimization problem.
For example, one can write $\bX=\bw\bw^H$ (rank-one) and minimize
$||\bw||^2=\trace\{\bX\}$ subject to constraints of the form $|\bw^H
\bg_i|^2 = \trace\{ \bX\bg_i \bg^H_i \} > \mbox{threshold}$ for all
$i$ by relaxing the rank constraint on $\bX$ into a semidefiniteness
constraint; once $\bX$ is found, keep its principal eigendirection.  A
next question is how to obtain \emph{multiple beamvectors} $\{\bw_k\}$
such that $\min_i \max_k | \bg_i^H \bw_k |$ is large.  (If the device
has a large enough supercapacitor, the relevant objective would
instead be $\min_i \sum_k | \bg_i^H \bw_k |$.)  Some initial results
on data-driven algorithms for this are given in \cite{thoota2023data}.

{\bf ii) Beamforming with direct-link interference suppression:}
Algorithms must be developed that can obtain (partial) channel state
information to the device and beamform towards it, at the same time as
direct link interference is suppressed. For example, for the bistatic
distributed MIMO setup, one can transmit into the nullspace of an
estimated channel between the transmitting and receiving panels
\cite{kaplan2023direct}, thereby reducing direct link interference.

{\bf iii) Panel selection and integration into the TDD flow:} For
communication with a backscattering device, a subset of [antenna]
panels needs be designated as transmitters and another subset as
receivers \cite{KaplanOL2024}. One approach is to quantify the prior
knowledge of the location (channel) of the backscattering device via a
probability distribution that can be learned and tracked over time,
and then find the panel assignment that maximizes the worst-case, or
average, performance over this distribution.  Furthermore,
communication with a backscattering device requires breaking the
nominal TDD flow, which calls for dynamic and flexible duplex access
operation patterns.

\subsection{ELAA with RIS}

Reconfigurable intelligent surfaces (RIS) are surfaces which can be electronically controlled to change their electro-magnetic reflection properties \cite{Dardari20}. Thereby, the channel properties can be adapted to the requirements of the communication system, e.g., to improve the scaling with the number of antennas, reduce the transmit power with number of RIS elements \cite{Zhi22}, enlarge the coverage region \cite{SangWCM,SangTWC}, and etc. 

The main difference of the RIS-assisted signal model to the one presented in Section III is that the new channel matrix $\widetilde{\Hb}_i$ in (\ref{eq:1}) contains the contributions of the direct link $\Hb_i$ and the contributions from the reflection of the RIS $\Gb \Theta \Fb_i$ with channel matrix $\Fb_i$ from the $i$-DN to the RIS and channel matrix $\Gb$ from the RIS to the user. The matrix $\Theta$ describes the operation at the RIS and its structure and its constraints depend on the RIS technology applied. The overall channel from the DN $i$ to one user is given by
\begin{eqnarray}
    \widetilde{\Hb}_i = \Hb_i + \Gb \Theta \Fb_i \label{eq:eCRIS}.
\end{eqnarray}
The classical constraint sets of the RIS matrix $\Theta$ is the diagonal matrix with unit modulus entries corresponding to phase shifts of each RIS element. 
Under this constraint, the design of RIS assisted multi-cell transmission systems is considered in \cite{li2023ris,jiang2022joint}. To support the connectivity of users in blind areas while ease the CSI acquisition overhead at each DN, statistical CSI based transmission design is a reasonable choice. With only statistical CSI at both the DNs and the RIS, it is obtained that the statistical maximum ratio transmission beamforming is almost optimal at each DN. This allows each DN to design the beamforming vectors of their own independently, which greatly reduces the design complexity. Moreover, the RIS phase shift design is also decoupled with the DNs' beamforming design. Besides the unit modulus constraint, discretized diagonal RIS models exists, where only a certain number of equidistant phases are supported \cite{SangTVT, Chenpeng23WCL}. More recently, the beyond-diagonal RIS model is introduced, where circuitry allows to transfer received signals between RIS elements \cite{Li23}. The constraint set in this case correspond to symmetric and unitary matrices 
\begin{eqnarray}
    \Theta = \Theta^T \quad \text{and} \quad \Theta^H \Theta = \bI \label{eq:RIScon}.
\end{eqnarray}
In the single-user scenario for one DNs, for fixed transmit $\wb_i$ and receive beamforming $\vb$, the resulting RIS optimization problem can be cast as 
\begin{eqnarray}
    \max_{\Theta \; \text{fulfills} \; (\ref{eq:RIScon})} \quad |\vb^H \Hb_i \wb_i + \vb^H \Gb \Theta \Fb_i \wb_i|^2 \label{eq:maxsuris},
\end{eqnarray}
which can be solved in closed form by exploiting the Takagi factorization as explained in \cite{Santamaria2023}. In the case with multiple DNs, the optimization problem is more difficult 
\begin{eqnarray}
    \max_{\Theta \; \text{fulfills} \; (\ref{eq:RIScon})} \quad \sum_{i=1}^C | \vb^H (\Hb_i + \Gb \Theta \Fb_i) \wb_i|^2. \label{eq:optSumRIS}
\end{eqnarray}
The programming problem in (\ref{eq:optSumRIS}) is difficult because the sum cannot be easily decomposed and $\Theta$ cannot be separated. 

For the multi-user MIMO case, the effective channel for user $k$ in the downlink from $C$ DNs can be written as 
\begin{eqnarray}
    \widetilde{\Hb}_{i,k} = \sum_{i=1}^C \left( \Hb_{i,k} + \Gb_k \Theta \Fb_i \right), 
\end{eqnarray}
and the achievable rate expression for user $k$ reads as in (\ref{eq:ardw}) with the difference that the effective channels contain contributions from the RIS. 
For the classical setup of MCP precoding with RIS optimization, the model and programming problem is formulated in \cite{soleymani2022rate}. 

The joint optimization of transmit covariance matrices and RIS-elements is performed in combination with next generation multiple access technologies, namely NOMA in \cite{soleymani2023noma} and rate splitting multiple access (RSMA) in \cite{soleymani2023optimization}. For simultaneous transmission and reflection (STAR) RIS configuration, the corresponding model and problem are formulated and solved in \cite{soleymani2023spectral}.

\begin{figure*}
\centering
\includegraphics[scale=0.42]{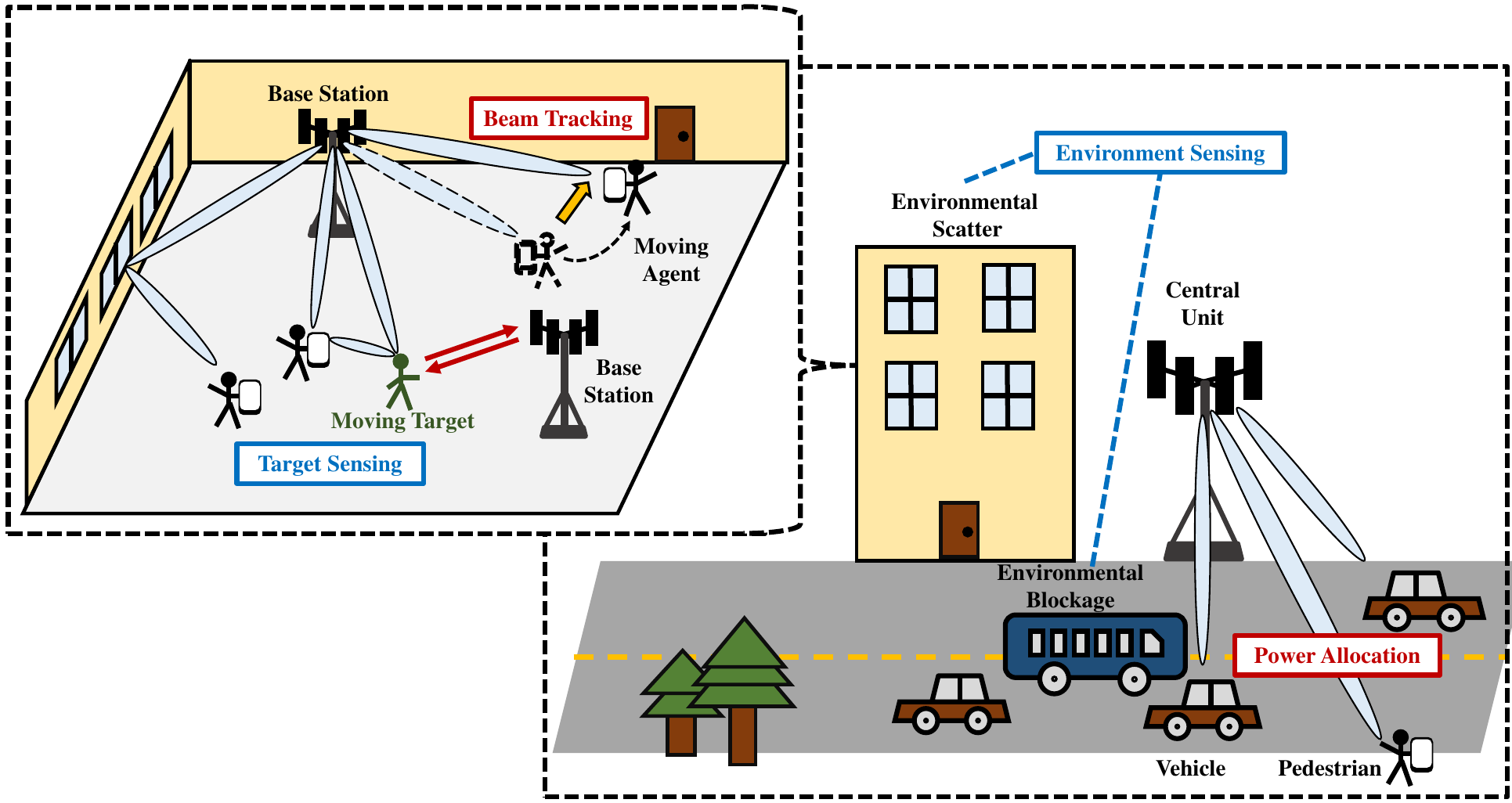}
\captionsetup{justification=justified, singlelinecheck=false, font=small}
\caption{ELAA-enabled ISAC functions in distributed systems.}
\label{f1}
\end{figure*}

In \cite{Chien2022}, RIS-assisted CF-mMIMO transmission over spatially correlated channels is optimized. An asymptotic analysis of the performance for growing number of APs and RIS elements is performed. The performance benefits of using RISs in CF-mMIMO systems are confirmed.

There are several open questions and topics for future research. First of all, the optimization of the RIS elements for different constraint sets including diagonal, BD-, STAR-RIS, globally passive, partly active under distributed massive MIMO has open challenges. If discrete constraint sets for the RIS elements are considered, tools and methods from discrete optimization need to be applied. 

Second, in many scenarios the RIS should be placed either close to the transmitter or to the receiver, if it is totally passive \cite{Pei21}. Then, it will operate most likely in the near-field of the corresponding transmitter or receiver. The resulting effect channel is then a quintuple $\Rb \Phi \Hb \Theta \Tb$, where the channel from the transmitter to the first RIS is denoted by $\Tb$, the channel from first RIS to second RIS by $\Hb$, and the channel from the receiver RIS to the receiver by $\Rb$. The two RIS matrices are denoted by $\Phi$ and $\Theta$. A joint optimization of $\Phi$ and $\Theta$ remains an interesting problem.

\subsection{ELAA with ISAC}

The future of ISAC systems will witness a profound fusion of communication and sensing capabilities, evolving from hardware, spectrum, and functional integration to a state of mutual benefit, high coordination, and deep integration \cite{Liu22}. 6G networks transcend the limitations of single-BS and single-UE perception, adopting an inherent sensing design across network architecture, networking technologies, and air interface capabilities to achieve high-precision perception.
Leveraging the 6G network, collaborative sensing technologies will emerge as a key component of ISAC system. This technology enhances the user experience at the edge and enables high-resolution sensing of multiple targets \cite{zhang2024optimal,strinati2024towards,guo2024integrated}. Collaborative sensing involves the cooperation and interaction of nodes within the mobile communication network, including large-scale deployed BS nodes and densely distributed UE nodes. By utilizing multiple nodes for receiving and transmitting sensing signals, collaborative sensing avoids the self-interference issues inherent in single-node perception, reducing the demand for receiver self-interference cancellation capabilities and lowering hardware costs.

ELAA systems are inherently suited for ISAC due to their shared hardware resources and similar signal processing methods. As for sensing, the same antenna array used for transmitting and receiving communication signals can also be utilized for sensing by analyzing the reflected echoes or the communication pilots. ELAA-enabled ISAC unlocks significant advantages in terms of performance. The ability of ELAA to realize precise beamforming translates to increased spatial resolution and effective range for sensing services, which brings about high-resolution UE localization and high-resolution environmental map construction \cite{Yang23,Que23}. 

For communication, ELAA-enabled ISAC enables highly targeted data transmission with minimal interference in multiple UE scenarios.  The spatial awareness offered by ELAA sensing capabilities can be used to optimize resource allocation, scheduling, and interference management across multiple UEs and BSs. For instance, by sensing the movements of multiple UEs and the environmental maps \cite{Yang22,Kim20}, an ELAA-enabled ISAC system can dynamically adjust beamforming patterns and allocate resources to maximize throughput and minimize interference. Similarly, in a multi-BS setup \cite{Du23}, sensing data can facilitate coordinated beamforming and interference mitigation strategies among BSs, leading to a more efficient and robust network. Furthermore, joint signal processing techniques can leverage information from both sensing and communication tasks to improve overall system performance by optimizing metrics of joint design. The ISAC functions in ELAA systems are shown in Fig. \ref{f1}, with sensing functions in blue, and communication functions in red.

Despite the considerable potential, several challenges remain in realizing the full potential of ELAA-enabled ISAC. Efficient power allocation and interference management are crucial to ensure both functionalities operate optimally without compromising each other. The design of sophisticated joint signal processing algorithms that effectively exploit the synergy between sensing and communication data is another significant challenge. Additionally, the hardware complexity and associated costs of implementing ELAA systems with a large number of antenna elements can be substantial and require further research and development for cost-effective solutions. 
ELAA-enabled ISAC in distributed systems holds immense promise for revolutionizing various fields. By addressing the remaining challenges and continuing to innovate in this domain, we can unlock the full potential of this transformative technology and create a future where sensing and communication seamlessly coexist and augment each other.

\section{Outlook for Future Directions} \label{sec: outlook}
In this section, we outline several important future research directions that are essential for improving the performance and practicality of ELAA systems.

\subsection{Distributed SP for ELAA in the Near Field}

In addition to fronthaul cost and computational complexity bottlenecks, ELAA systems face unique challenges due to the extremely large number of antennas operating in the radiating near-field region. Unlike conventional wireless systems functioning in the far-field, near-field communications fundamentally alter the electromagnetic field properties. Two key phenomena emerge in this context: spatial nonstationarity and spherical wavefronts. Spatial nonstationarity means different parts of the antenna array may experience the same channel paths with varying power or entirely different channel paths, while spherical wavefronts account for the curvature of signal propagation due to closer proximity between transmitter and receiver \cite{bjornson2020power,Amiri2018extremely}.

These phenomena present both challenges and opportunities for signal processing. For instance, near-field effects complicate channel estimation and beamforming but also offer enhanced spatial degrees-of-freedom and improved interference mitigation capabilities. Existing studies have explored these complexities: spatial nonstationarity-focused channel estimation has been examined in \cite{tan2023threshold}, spherical wavefront channel estimation in \cite{cui2022channel}, and joint investigations of both phenomena in \cite{han2020channel}. Additionally, \cite{zhang2022beam,ding2023noma,ding2023resolution} demonstrated that precise downlink beamforming leveraging these effects could enable focused beams, providing a novel degree of freedom to reduce multiuser interference.

To fully exploit the potential of near-field ELAA systems, innovative distributed SP algorithms are necessary. Unlike in far-field systems, where DBP architectures rely on preconfigured antenna clusters, near-field spatial nonstationarity calls for dynamic clustering approaches. Such methods could optimize antenna utilization by identifying and leveraging sparsity patterns in the antenna-domain channel. Furthermore, spherical wavefronts present new opportunities for enhancing both communication and localization, necessitating distributed algorithms that are specifically designed to account for curvature effects in wave propagation.

Finally, while addressing these technical issues, it is critical to consider practical deployment scenarios. The susceptibility of near-field ELAA systems to interference from nearby scatterers and the challenges of maintaining synchronization among distributed nodes further emphasize the importance of robust and efficient algorithmic designs. Future research should prioritize developing SP algorithms that balance these challenges with the unprecedented advantages of near-field ELAA systems.

\subsection{ELAA Realizations with Flexible Antennas}
The development of ELAA systems is poised to benefit significantly from advancements in antenna technologies. Incorporating novel antenna designs can further enhance the performance and flexibility of ELAA systems. Recently, several advanced antenna techniques have emerged, such as fluid antennas \cite{wong2022bruce} and movable antennas \cite{zhu2023movable}. Unlike conventional antenna techniques, where antennas are fixed in position, these new technologies allow for flexible adjustment of antenna positions within a spatial region \cite{wong2020fluid,zhu2023modeling,ma2023mimo,wong2020performance,shao20246dma}.
With flexible antenna positions, the system can dynamically adapt to changing environments, providing more consistent channel conditions to enhance the received signal power. However, in these flexible-antenna systems, the movement of antennas is typically constrained to the scale of the wavelength, which limits their impact on large-scale path loss and restricts their applicability in various scenarios. 

More recently, the pinching antenna has emerged as a more promising solution to address the limitations of conventional flexible antenna systems \cite{suzuki2022pinching,ding2024flexible}. Unlike the other flexible antenna systems, pinching antennas utilize a dielectric waveguide as the transmission medium, allowing for dynamic activation at any point along the waveguide by adding or removing separate dielectric materials. This innovation enables much greater flexibility in antenna deployment. Additionally, compared to traditional flexible antennas, pinching antennas are more cost-effective and simpler to install, as their functionality is achieved through the straightforward process of adding or removing dielectric materials. These advantages make pinching antennas particularly well-suited for environments where adaptability and cost-efficiency are key, such as in industrial Internet of Things (IoT) applications or urban deployments.

While these new flexible antenna technologies offer significant promise, they also introduce new research challenges. Developing efficient algorithms to control and optimize flexible antennas, ensuring seamless integration with existing network infrastructure, and addressing potential cost and complexity issues are critical areas for future research. Additionally, investigating the impact of these technologies on network performance, user experience, and overall system reliability will be essential to realize their full potential.

\subsection{ELAA for Physical Layer Security}
Physical layer security (PhySec) refers to algorithms and schemes which exploit physical parameters and properties of the transceiver as well as the wireless channel to obtain security primitives. This is considered as the first line of defense for $6$G wireless communication networks \cite{Mucchi2021}. 

ELAA can support PhySec schemes, like, e.g., wiretap coding, secret key generation, and authentication. Due to the distributed large number of transmit antennas, the beamforming can focus the energy towards the intended receivers while reducing information leakage to eavesdroppers. The remaining information leakage is handled by wiretap coding. Following the signal model for the ELAA downlink precoding from (\ref{eq:dpelaa}) for the case with one sub-carrier, the sum secrecy rate maximization problem for confidential data transmission reads as follows
\begin{eqnarray}
    \max_{ \{\Zb_{i,k} \}} \sum_{k=1}^K SR_{k} \quad \text{s.t.} \quad \eqref{eq:11b},
\end{eqnarray}
where $SR_k = [R_k - R_\ell]^+$ is the achievable secrecy rate. It is equal to the achievable rate of user $k$ minus the achievable rate to the eavesdropper $\ell$. While for single-user, the global optimal precoding is known \cite{Mukherjee22}, for the multi-user case only algorithms to achieve the stationary solution exist \cite{Choi21}. For CF-mMIMO, the physical layer secrecy is optimized for downlink transmission in \cite{Tubail2023}. Very recently, PhySec with near-filed beamforming is instroduced in \cite{Ferreira24}. For secret key generation (SKG) in multi-carrier MIMO networks exists efficient and robust solutions \cite{Li21} also against various attacks. SKG in downlink massive MIMO is studied in \cite{Li21a}. However, SKG algorithms that takes the constraints and properties of the ELAA systems into account are not available yet. 

The wireless fingerprints, which are essentially the measured effective channels between the transmitter and receiver \cite{Xie21}, contain more information as the number of antennas  increases \cite{Maeng22}. For the CF-mMIMO architecture, \cite{Qiu22} studies user location estimation method based on fingerprint positioning. Important aspects are the careful selection of the AP and corresponding local channel estimation.

\subsection{Low-Resolution Designs for ELAA Systems}
Recent research on multi-antenna systems has shown substantial interest in techniques related to low-resolution or coarsely quantized signals from a fundamental signal processing perspective \cite{jacobsson2015one,jacobsson2017throughput,wang2018hybrid,li2017channel,shao2019one}. This trend is driven by the potential to replace high-resolution analog-to-digital converters (ADCs) and digital-to-analog converters (DACs) with lower resolution alternatives, particularly the very cheap one-bit ADCs/DACs \cite{li2017channel,shao2019one}. Utilizing such low-resolution schemes can significantly reduce the hardware complexity and power consumption of radio-frequency front ends. For ELAA systems, low-resolution designs become even more critical because the number of ADCs/DACs and radio-frequency front ends needs to scale proportionally with the very large number of antennas. Consequently, addressing issues related to hardware cost and energy consumption is crucial, as these could otherwise become prohibitively expensive in ELAA systems.

There have been numerous studies exploring low-resolution designs in massive MIMO systems. For example, \cite{li2017channel,fesl2024channel,atzeni2021channel} investigated the one-bit channel estimation problem in massive MIMO systems. The impact of coarse quantization on channel estimation accuracy and achievable rate was studied in \cite{li2017channel}, and efficient channel estimation algorithms were developed based on Gaussian latent models in \cite{fesl2024channel}. One-bit precoding designs, using criteria such as MMSE and minimum symbol error probability, have been explored in \cite{sohrabi2018one,shao2019framework}. Another important challenge is uplink MIMO detection with low-resolution ADCs/DACs. The research in \cite{risi2014massive} explored the application of linear receivers in the low-resolution scenario, while maximum likelihood-based detectors were investigated in \cite{studer2016quantized,shao2020binary,wen2015bayes}.

While these low-resolution schemes have demonstrated promising numerical performance, there are significant challenges. The primary difficulty lies in the development of efficient non-convex algorithms that can handle large-scale binary optimization problems. Such algorithms are critical for making ELAA systems practical, particularly when addressing the need for high-performance while keeping hardware and computational costs low. Additionally, low-resolution designs inherently introduce trade-offs, particularly in terms of channel estimation accuracy, signal-to-noise ratio (SNR), and system capacity. Although they can reduce hardware complexity, they also tend to degrade system performance due to quantization errors. This makes it crucial to find a balance between cost savings and performance loss. Future research should focus on strategies to mitigate these trade-offs, possibly by developing hybrid ADC/DAC systems or advanced signal processing techniques that minimize the impact of quantization errors.

\subsection{ELAA for Satellite Communications}
Satellite communications (SATCOM) are expected to play a pivotal role in the future wireless ecosystem, especially in providing global coverage for remote and underserved areas. As demand for high-speed, low-latency communication continues to rise across industries like the Internet of Things (IoT), autonomous systems, and disaster recovery, SATCOM must evolve to meet the demands of the 6G era \cite{chen2020system}. Recent advances have demonstrated that the application of ELAA (in the form of distributed MIMO) technologies to SATCOM is emerging as a promising solution for improving system performance, capacity, and coverage \cite{he2024spatial,abdelsadek2022distributed}.

The use of ELAA systems in SATCOM offers several unique advantages. By deploying large numbers of antennas on both the satellite and ground stations, ELAA systems can significantly enhance beamforming precision, thereby mitigating interference, improving signal coverage, and increasing capacity, especially in high-user-density environments or areas with challenging propagation conditions. However, the deployment of ELAA systems in SATCOM also introduces substantial challenges. The large size and complexity of antennas required for ELAA can increase costs and pose challenges in satellite payload design. Additionally, the need for precise beamforming and the handling of large-scale antennas in space introduces significant technical complexity \cite{xiao2022leo}. Furthermore, the integration of ELAA systems with terrestrial networks and overcoming the inherent latency of satellite communications require advances in distributed SP algorithms to ensure seamless and efficient operation.

Despite these challenges, ELAA for SATCOM presents significant potential to enhance system performance and is an important research area for the next-generation communication systems. Addressing the technical challenges associated with large-scale antenna integration, beamforming, and distributed SP will be critical for realizing the full potential of ELAA in SATCOM for next-generation wireless communications.

\section{Conclusion} \label{sec: conclusion}

ELAA systems are crucial for realizing the full potential of 6G
wireless communication networks. As the number of antennas in ELAA
systems increases, significant challenges such as excessive
interconnection costs and high computational complexity
emerge. Efficient distributed SP algorithms have shown great promise
in addressing these challenges.  We began by presenting three
representative forms of ELAA systems: single-BS ELAA systems, coordinated
distributed antenna systems, and ELAA systems integrated with emerging
technologies. For each form, we reviewed the associated distributed SP
algorithms designed to mitigate the bottlenecks.  Additionally, we
highlighted several important future research directions, including
distributed SP algorithm design in the near-field, low-resolution
design for ELAA systems, ELAA for physical layer security, and the
incorporation of advanced antenna technologies. Addressing these areas
is essential for further improving the performance and practicality of
ELAA systems.  In conclusion, while significant progress has been made
in the development of distributed SP algorithms for ELAA systems,
ongoing research and innovation are still required to overcome
existing challenges and fully realize the potential of ELAA in
next-generation wireless networks.


\smaller[1]

\end{document}